%% file: main.tex
\documentclass[twocolumn, twocolappendix]{aastex701}

\usepackage[T1]{fontenc}
\usepackage{amsmath}
\usepackage{xspace}
 
\graphicspath{{figures/}}
\input{include/def}

\shorttitle{Bursty star formation and spectral diversity at high-$z$}
\shortauthors{Gelli et al.}
 
\begin{document}
 
\title{How bursty star formation shapes spectral diversity at high redshifts:\\breaking degeneracies in early galaxy formation}
 
\author[0000-0001-5487-0392]{Viola Gelli}
\affiliation{Cosmic Dawn Center (DAWN), Copenhagen, Denmark}
\affiliation{Niels Bohr Institute, University of Copenhagen, Jagtvej 128, 2200 K{\o}benhavn N, Denmark}
\email[show]{viola.gelli@nbi.ku.dk}
\correspondingauthor{Viola Gelli}

\author[0000-0002-3407-1785]{Charlotte A. Mason}
\affiliation{Cosmic Dawn Center (DAWN), Copenhagen, Denmark}
\affiliation{Niels Bohr Institute, University of Copenhagen, Jagtvej 128, 2200 K{\o}benhavn N, Denmark}
\email{charlotte.mason@nbi.ku.dk}
 
\author[0000-0001-6106-5172]{Daniel P. Stark}
\affiliation{Department of Astronomy, University of California, Berkeley, Berkeley, CA 94720, USA}
\email{dpstark@berkeley.edu}

\author[0000-0001-5940-338X]{Mengtao Tang}
\affiliation{Tsung-Dao Lee Institute, Shanghai Jiao Tong University, 1 Lisuo Road, Shanghai 201210, People’s Republic of China}
\affiliation{School of Physics and Astronomy, Shanghai Jiao Tong University, 800 Dongchuan Road, Shanghai 200240, People’s Republic of China}
\affiliation{State Key Laboratory of Dark Matter Physics, Shanghai Jiao Tong University, 1 Lisuo Road, Shanghai 201210, People’s Republic of China}
\email{mengtao.tang@sjtu.edu.cn}

\author[0000-0002-8984-0465]{Julian B. Muñoz}
\affiliation{Department of Astronomy, The University of Texas at Austin, Austin, TX, USA}
\affiliation{Cosmic Frontier Center, The University of Texas at Austin, Austin, TX 78712}
\email{julianbmunoz@utexas.edu}

\author[0000-0002-2178-5471]{Zuyi Chen}
\affiliation{Cosmic Dawn Center (DAWN), Copenhagen, Denmark}
\affiliation{Niels Bohr Institute, University of Copenhagen, Jagtvej 128, 2200 K{\o}benhavn N, Denmark}
\email{zuyi.chen@nbi.ku.dk}

\author[0000-0001-9125-7415]{Ivan Nikolić}
\affiliation{Cosmic Dawn Center (DAWN), Copenhagen, Denmark}
\affiliation{Niels Bohr Institute, University of Copenhagen, Jagtvej 128, 2200 K{\o}benhavn N, Denmark}
\email{ivan.nikolic@nbi.ku.dk}

\begin{abstract}
JWST has revealed an unexpected abundance of $z>10$ galaxies and striking spectral diversity.
We model bursty star-formation histories (SFHs) to explore how they shape this diversity and whether spectroscopy can discriminate between explanations for the UV luminosity function (UVLF) excess.
We find strong burstiness ($\sigma_{\rm UV}(M_h=10^{10}\,M_\odot)\sim0.9$\,mag), increasing toward low-mass halos, is required to reproduce the $z\sim6$ UVLF and observed fractions of galaxies in recent star-formation upturns and downturns, traced by $\rm SFR_3/SFR_{50}$.
This burstiness naturally reproduces UVLFs and upturn and downturn fractions over $z\sim6-11$.
With increasing redshift, we predict galaxies in recent upturns ($\rm SFR_3/SFR_{50}>5$) increasingly dominate the UVLF bright-end, reflecting rising accretion rates and burstier low-mass halos. 
This results in higher median emission-line equivalent widths (EWs), bluer UV-slopes and weaker Balmer breaks at fixed $M_{\rm UV}$, while distributions of these properties broaden.
We find the downturn fraction ($\rm SFR_3/SFR_{50}<0.2$) is particularly sensitive to stochasticity: 
increasing towards faint $M_{\rm UV}$ if burstiness increases in lower-mass halos, and evolving with redshift only if stochasticity does.
However, the model underpredicts the $z\sim14$ UVLF, so we explore two models which match the UVLF: \textit{increased burstiness}
and \textit{increased star-formation efficiency}.
From $z=10$ to 14, at fixed $M_{\rm UV}$, increased burstiness predicts the downturn fraction increases by $\sim2\times$ and the weak-emitter fraction (H$\beta$\,EW$\,<50\,\angstrom$) by $\sim1.5\times$. 
By contrast, the increased-efficiency model predicts a constant downturn fraction and declining weak-emitter fraction.
Overall, our results indicate burstiness is important for explaining spectral diversity and evolution at $z\sim6-14$, and that spectroscopic samples spanning a broad $M_{\rm UV}$ range can establish its role in the $z\sim14$ UVLF excess.

\end{abstract}

\keywords{\uat{Galaxy formation}{595} --- \uat{High-redshift galaxies}{734}}
 
\section{Introduction}\label{Sec:intro}

Among the most striking discoveries from the James Webb Space Telescope (JWST) are the numerous luminous galaxies at $z>10$ revealed by early NIRCam campaigns \citep[e.g.][]{Castellano22, Naidu22schrodinger, Adams2022, Bouwens23, Donnan2022, McLeod24, Robertson2024, Whitler25}. In the past two years, the discovery and subsequent spectroscopic confirmation of galaxies at $z\gtrsim14$ \citep[e.g.][]{Carniani2024, Naidu2026_MoM} has pushed the observational frontier to only $\lesssim300$ Myr after the Big Bang.
The high abundance of $z>10$ galaxies poses one of the most significant challenges to current models of galaxy formation. Standard models developed before JWST, tracing the evolution of the halo mass function, predicted a much steeper decline in the number density of luminous galaxies toward these early epochs \citep[e.g.,][]{Mason15, tacchella:2018, Yung2020}. Even new models specifically tuned to match the observed galaxy population at $z\sim10$ still struggle to reproduce the latest UVLF constraints at $z\sim14$ \citep[e.g.][]{Weibel2025, Whitler25, Chemerynska2026}.

The majority of explanations proposed to explain the UVLF fall into two classes:
either galaxies are intrinsically brighter at fixed halo mass relative to pre-JWST expectations; or the scatter in UV luminosity increases, allowing galaxies hosted by lower-mass halos to temporarily populate the bright end of the UV luminosity function.
The first case includes scenarios with increased star formation efficiency (SFE) due to inefficient feedback \citep{Dekel23, Li23, Chen2026_FFB} or efficient cloud scale star formation \citep[e.g.][]{Menon2024, Somerville2025}, a top-heavy initial mass function 
\citep[e.g.][]{Inayoshi22, Rasmussen23, Trinca23, Hutter2025}, or reduced dust attenuation due to radiation pressure \citep[e.g.][]{Ferrara23, Fiore23} or changes in the grain size distribution \citep[e.g.][]{McKinney2025, Narayanan2025, Sommovigo2026}.
The second class invokes bursty star formation histories, where feedback driven SFR fluctuations can temporarily boost UV luminosities in the numerous low-mass halos at high redshift into the bright end of the UV luminosity function \citep[e.g.][]{Mason2023a, Pallottini23, Shen23, Sun23b, Kravtsov24, Nikolic24}. 
Recent models including halo mass dependent burstiness \citep[e.g.][]{Gelli24b, Munoz2026}, motivated by predictions of feedback becoming increasingly efficient in lower-mass systems with shallow potential wells \citep[e.g.][]{Faucher-Giguere2018, Ma18, Gelli20, Furlanetto22}, showed that this can further boost the UV luminosity function, bringing models predictions into good agreement with JWST measurements up to $z\sim12$.
However, reproducing the observed UVLF at $z\sim14$ is still a major challenge. Crucially, models with increased SFE and increased burstiness both enhance the number densities of galaxies at fixed $M_{\rm UV}$, making it impossible to distinguish between these scenarios with currently observed luminosity functions alone.

Spectroscopy offers a promising way to break this degeneracy. Different spectral features are sensitive to star formation on different timescales, providing direct constraints on the underlying star formation histories (SFH) and its burstiness.
As the first JWST NIRSpec prism spectra of luminous $z>9$ galaxies were taken, they revealed a remarkable diversity in their physical properties.
Some UV-luminous systems exhibit extremely strong UV emission lines, potentially linked to their compact morphologies and extreme surface brightness densities \citep[e.g., GN-z11, GHZ2, MoM-z14][]{Bunker2023, Castellano2024, Naidu2026_MoM}. 
These galaxies are often interpreted as compact starbursts undergoing rapid recent increases in star formation, or possibly hosting active galactic nuclei \citep[e.g.][]{Harikane2025, Kokorev2025}. In contrast, other luminous galaxies show weak emission lines and more extended morphologies \citep[e.g., Gz9p4, GS-z14][]{Boyett2023, Carniani2024, Chen2026}. 
While the origin of weak emission line systems has been challenging to interpret in low resolution prism spectra, recent studies with additional constraints from ultra-deep NIRSpec grating spectroscopy \citep[e.g.][]{Chen2026} and MIRI follow-up to access rest-frame optical nebular emission lines and Balmer breaks at $z>10$ \citep[e.g.][]{Helton2025, Harikane26} suggest that, at least some, UV-bright weak-emission line galaxies may be observed after a recent downturn phase in their star formation histories.
However, while detailed studies of individual objects are critical for understanding the physical conditions in early galaxies, they provide only single snapshots of star formation histories. Progress in understanding the physical origin behind the $z\sim14$ UVLF excess now requires large statistical samples to characterize the full galaxy population properties.

In the low redshift universe, spectral diagnostics such as combination of UV and Balmer-lines have long been used to constrain burstiness and SFH in large spectroscopic samples \citep[e.g.][]{Weisz12, Guo2016, Emami2019}. JWST has now made the same approach possible at $z\sim6$, where observations reveal large scatter in the star-forming main sequence \citep[e.g.][]{Cole23, Ciesla23, Simmonds2024, DiCesare2026, Stoffers26}, broad distributions of emission-line equivalent widths at fixed UV luminosity \citep[e.g.][]{Endsley23, Navarro-Carrera2026}, and substantial diversity in inferred ionizing photon production efficiencies \citep[e.g.][]{PrietoLyon2023a, Simmonds24_ion}. These broad distributions are naturally expected if galaxies are observed at different phases of highly time-variable SFHs, including intense starbursts, downturn phases, and temporary quenching episodes following feedback-driven suppression of star formation \citep[e.g.][]{Looser23b, Strait23, Looser23, Trussler24, Baker25}.

We are now at a stage where these analyses can be extended to $z>9$. 
Thanks to on-going spectroscopic efforts, particularly with NIRSpec prism (e.g. JADES, CEERS, RUBIES, CAPERS; \citealt{Eisenstein2023a, Bunker2023b, Finkelstein23, deGraaff2025_rubies, Donnan2025_capers}),
JWST is now delivering increasingly large spectroscopic samples of bright galaxies at $z\gtrsim9$.
These samples are enabling statistical studies and revealing a huge variety in galaxies' spectral properties, including broad distributions of UV slopes and Balmer-line equivalent widths \citep[EW, e.g.][]{Roberts-Borsani2024, Saxena2024, Topping2024, Tang2025, Roberts-Borsani2026}, suggesting highly stochastic SFHs in the early universe.
These spectra reveal not only significant diversity, but also evidence of evolution of spectral properties with redshift at fixed UV luminosity. For example, \citet{Tang2025} find that the fraction of galaxies with strong emission lines increases from $z\sim6$ to $z\sim10$. Understanding and interpreting these trends requires a framework that connects star formation histories to the observed statistical distributions of spectral diagnostics and predicts their evolution toward $z\sim14$.

In this work, we develop such a framework by combining bursty star formation histories modeling with spectral synthesis. 
We create a dark matter halo-based star formation history model, calibrated at $z\sim6$, and explore its evolution at $z\sim10-14$.
We test whether different scenarios for high-$z$ star formation histories that reproduce the $z\sim14$ UV luminosity function, such as enhanced star formation efficiency and increased burstiness, predict distinct distributions of spectral diagnostics, providing a way to identify the physical origin of the unexpectedly abundant luminous galaxies at the highest redshifts.

The paper is structured as follows: in Section~\ref{Sec:methods} we describe the model for populations of galaxies with bursty star formation histories; in Section~\ref{Sec:fiducial} we calibrate and anchor the model based on observations at $z=6$; in Section~\ref{sec:z6_10_evol} we explore the predicted evolution from $z=6$ to 10; in Section~\ref{sec:highz} we explore different physical scenarios at $z=14$ and their impact on a number of observables distributions; in Section~\ref{sec:discussion} and ~\ref{sec:conclusions} we present a discussion of the results and summary of the main conclusions.

Throughout the paper we assume a flat $\Lambda$CDM  cosmology with $\Omega_m=0.3$, $\Omega_\Lambda=0.7$ and $h=0.7$.

\section{Methods}\label{Sec:methods}

We aim to build a framework to model bursty star formation histories and emission for large populations of high-redshift galaxies.
In the following, we describe how we model the time evolution of star formation in individual halos, then how we compute their spectral energy distributions (SED) to predict observable properties, and finally how we construct galaxy populations across different redshifts.

\subsection{Halo growth and baseline star formation rate}

We first derive the mass-assembly history of each dark-matter halo. Starting from its mass at a given redshift, we reconstruct its growth history using an analytic halo-growth model. The halo growth follows the Einstein–de Sitter high-redshift approximation\footnote{We tested the impact of this assumption by using a different accretion-rate parametrization from \citet{Yung2024_gureft}, based on N-body dark-matter-only simulations of the early Universe, and found no change in our conclusions.}, as derived in e.g. \cite{Dekel23}:

\begin{equation}
    \frac{\dot{M_h}}{M_h} \simeq 0.03\,\mathrm{Gyr}^{-1} \left( \frac{M_h}{10^{12}M_\odot} \right)^{0.14}(1 + z)^{5/2}.
\end{equation}

We use this halo-growth history to define a baseline (smooth) star-formation history, onto which stochastic fluctuations will be added. We assume that the star formation rate traces the baryonic accretion rate onto the halo \citep[e.g.][]{Mason15, tacchella:2018, Shen23, Yung2025}:

\begin{equation}
    \mathrm{SFR}_{\rm baseline}(t) = \epsilon_\star(M_h) \,\, f_b  \,\, \frac{dM_h}{dt} \label{eq:sfr_t}
\end{equation}
where $f_b=\Omega_b/\Omega_m$ is the baryon fraction and $\epsilon_\star(M_h)$ the mass-dependent but redshift-independent instantaneous star formation efficiency. This efficiency controls how accreted baryons are converted into stars and is therefore a key parameter of the model. Following previous models \citep[e.g.][]{Moster2010, Shen23, Munoz2026}, we parameterize the star formation efficiency as a double power law:
\begin{equation}
    \epsilon_\star(M_h) = \frac{2 \epsilon_{\star,0}}{\,(M_h(t)/M_c)^\alpha +  (M_h(t)/M_c)^\beta\,}
    \label{eq:sfe}
\end{equation}
where we fix $\alpha = -0.6$, $\beta = 0.5$, $M_c = 10^{11.8}\msun$. These values correspond to the best-fitting $z\sim6$ star formation efficiency inferred by \cite{Munoz2026}, also consistent with observational constraints by \cite{Shuntov2025} and \cite{Harikane2022_LF_cluster}. 
The normalization of the star formation efficiency $\epsilon_{\star,0}$ is a free parameter of our model.

\subsection{Stochasticity}

Both simulations and observations point to a picture where the star formation histories of galaxies undergo significant fluctuations over time \citep[e.g.][]{FaucherGigure18, Ma18, Gelli20, Iyer2020, Tacchella20, Looser23b, Bhagwat24, Ciesla23, Endsley23b, Endsley25}. These fluctuations arise from the complex interplay between gas accretion, gas depletion, and feedback from radiation, massive stars, and supernovae, which can heat and eject gas from the galaxy. This interplay leads to alternating phases of enhanced and suppressed star formation, giving rise to bursty star formation histories.
Modeling this variability is essential for interpreting high-$z$ observations, as such fluctuations can lead to substantial variations in the observed spectral properties of galaxies over time \citep[e.g.][]{Sun23a}.

Simulations show that SFR are typically found to be distributed normally in log-SFR space \citep[e.g.][]{Feldmann2017, Caplar2019, Iyer2020, Pallottini23}.
To analytically describe such realistic bursty star formation histories, we introduce a stochastic component to the baseline smooth SFR of Equation~\ref{eq:sfr_t}. For each galaxy, we assume that the instantaneous SFR fluctuates around the median baseline value:

\begin{equation}
    \rm SFR = SFR_{baseline}\times 10^{\Delta}
\end{equation}

where $\Delta$ describes the SFR fluctuations in log space.
We note that, by adopting this parametrization in log-space, the burstiness of SFH is responsible for driving higher \textit{mean} SFR with respect to the baseline \textit{median} value, which will lead to higher average luminosities whenever stochasticity is present. In other words, the $\epsilon_\star$ parameter determines the \textit{median} SFR trend, while $\Delta$ determines the \textit{mean} SFR and fluctuations around the median.

We describe the time-variability $\Delta$ in the frequency domain through a power spectral density (PSD). Following previous works \citep[e.g.,][]{Caplar2019, Tacchella20, Iyer2024, Kravtsov24, Sun2025_burstyHa, Burnham2026}, we adopt a broken power-law PSD form:

\begin{equation}
    \mathrm{PSD}(f) = \frac{\sigma^2_{\rm \log SFR}(M_h)}{1 + (\tau f)^\alpha}
\end{equation}

where $f$ is frequency, $\sigma$ is the amplitude, $\tau$ is a characteristic timescale over which the fluctuations become uncorrelated. The parameter $\alpha$ sets the high-frequency slope of the PSD, such that larger values of $\alpha$ suppress short-timescale fluctuations more strongly, resulting in smoother star formation histories. Different theoretical models and simulations consistently find $\alpha=2$ \citep[][]{Caplar2019, Tacchella20}, which we fix throughout this work.

The characteristic timescale $\tau$ is a free parameter in our model and it is expected to be tightly related to the physical processes regulating star formation in galaxies. For instance, different feedback mechanisms typically suppress star formation on different characteristic timescales, e.g. Active Galactic Nuclei and radiation pressure on $\lesssim 10$~Myr as they act efficiently after their onset, supernovae on $\sim10-50$~Myr as they reflect the delay due to lifetimes of different stars \citep[e.g.][]{starburst99, Hopkins2014, Furlanetto22}, and mergers on $>100$~Myr \citep[e.g.][]{Boylan-Kolchin2008, Hani2020}.

The amplitude of the fluctuations $\sigma_{\rm \log SFR}$ is also expected to strongly depend on the galaxies' physical properties. Several models and simulations suggest a significant halo mass dependence: low-mass halos, with their shallow gravitational potential wells, are very sensitive to feedback processes and easily experience strong variations in the SFR, whereas more massive systems are typically more stable against both internal and external perturbations \citep[e.g.][]{FaucherGigure18, Ma18, Gelli20}. 

For this reason, we model the SFR fluctuations' amplitude as a mass-dependent quantity in the form:

\begin{equation}
    \sigma_{\rm \log SFR}(M_h(t)) = \sigma_{10} + \gamma \log_{10}\left( \frac{M_h(t)}{ 10^{10} M_\odot} \right),
    \label{eq:mass_dep}
\end{equation}

where the normalization $\sigma_{10}$ corresponds to the amplitude at $M_h = \rm 10^{10} M_\odot$ and is treated as a free parameter in our modeling. 
We fix the mass dependent slope to a negative value of $\gamma=-0.15$ to produce higher burstiness in lower mass halos, reflecting the increasing ability of more massive halos to retain gas against feedback processes. The value was set following constraints obtained by \cite{Munoz2026} when fitting $\gamma$ as a free parameter, and it is also consistent with the results of \cite{Nikolic24} and \cite{Basu2026}.
We set a minimum amplitude of $\sigma_{\log\rm SFR}=0.3$ to account for the level of burstiness expected from stochastic dark matter accretion rates alone \citep{Ren18}.

We note that, because the amplitude of SFR fluctuations depends on halo mass, the level of burstiness naturally evolves with time as the halo grows: while the baseline SFR increases in time with halo mass, the level of burstiness gradually decreases, such that an individual halo experiences larger SFR fluctuations at earlier times than at later times. 
While this effect is often neglected in similar models \citep[e.g.][]{Iyer2020, Burnham2026, Munoz2026}, which assume a time-independent fluctuation amplitude for a given halo, our model captures the evolution of burstiness within individual halos. As a result, galaxies experience systematically burstier star-formation histories at earlier times, potentially affecting observables sensitive to star formation on longer timescales, such as the UV luminosity and Balmer-break strength (see the next Section~\ref{subsec:meth:spectralmodeling}).

\subsection{Spectral modeling}\label{subsec:meth:spectralmodeling}
 
As we aim to infer the impact of bursty star formation histories into high-$z$ observables, a key ingredient of our framework is the forward modeling of the spectral energy distributions (SED) of galaxies.

We generate SEDs using \code{STARBURST99} \citep{starburst99} with \code{Padova} stellar evolutionary tracks \citep{Bertelli94}, computing both stellar and nebular continuum emission for each galaxy based on its SFH.
The nebular continuum and emission-line equivalent widths are computed assuming that ionizing photons are fully absorbed by the surrounding gas and adopting standard case-B recombination.
The SFHs are sampled in logarithmic timesteps, with the final $\sim50$ Myr resolved at a very fine resolution ($\Delta t \leq 1~\mathrm{Myr}$) to accurately capture the effects of short-term variability on emergent emission.
We adopt a standard Kroupa IMF, a double power-law with slopes of $1.3$ and $2.3$ in the ranges [0.1, 0.5]~$\msun$ and [0.5, 100]~$\msun$ respectively \citep{Kroupa01}. We discuss the potential implications of varying this assumption in Section~\ref{sec:discussion}.
We fix the stellar metallicity to $Z = 0.1~Z_\odot$, appropriate for the high-redshift systems considered \citep[e.g.][]{Curti23, Nakajima23, Sanders2024}. While exploring a full metal-enrichment history would be interesting and will be addressed in future work, metallicity variations are expected to be sub-dominant to SFH variability on the distributions of spectral properties we explore in this paper \citep[e.g.][]{Shivaei2018, Gelli23, Dome24, Endsley23b}.

To estimate the UV magnitude $M_{\rm UV}$, defined as the rest-frame absolute AB magnitude at $\rm 1500~\angstrom$, we add dust attenuation following \cite{Mason15} and adopting \cite{Meurer1999} attenuation law $A_\mathrm{UV} = 4.43 + 1.99\beta$, including the UV slope $\beta(M_{\rm UV}, z)$ empirically from observations by \citet{Bouwens2014}. 
We do not consider dust attenuation at $z>10$, as recent results suggest its effect to be negligible in very high-$z$ galaxies \citep{Cullen2024, Morales2024, Topping2024}.

\begin{figure}
    \centering
    \includegraphics[width=0.85\linewidth]{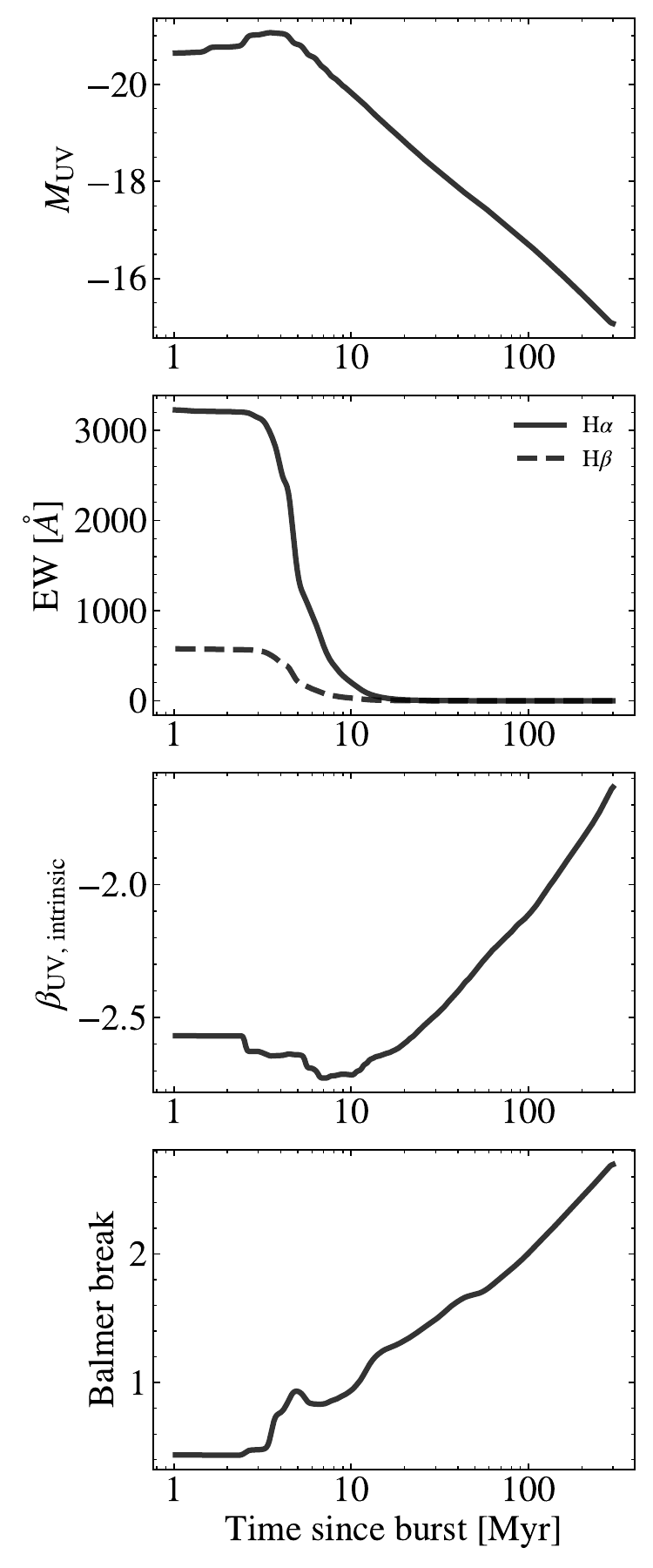}
    \caption{Evolution in time of UV magnitude, Balmer lines EW, UV slope, and Balmer break (defined as $F_\nu(4200\angstrom)/F_\nu(3500\angstrom)$) for a single burst of $M_\star=10^8\msun$ and metallicity of $Z~=~0.1\rm~Z_\odot$, using \code{starburst99} models \citep{starburst99}.}
    \label{fig:single_burst}
\end{figure}

From the modeled SED, we use a set of key observables, commonly used to characterize star formation in high-$z$ galaxies, that are sensitive to different timescales: UV magnitude, Balmer line equivalent widths, UV slope, and the strength of the Balmer break. To build intuition, in Fig.~\ref{fig:single_burst} we show the evolution of these observables, following a single burst of star formation (of $M_\star = 10^8\msun$) for 300~Myr.

The top panel shows the evolution of the intrinsic UV magnitude $M_{\rm UV}$, i.e. prior to dust attenuation. The UV emission peaks at $\sim 2-5$~Myr due to the post-ZAMS evolution of massive ($\gtrsim15\msun$) stars, including the onset of Wolf–Rayet phases, and subsequently gradually declines as these stars end their lives.
At later times ($>5$~Myr), the UV continuum still receives contribution from longer lived lower mass stars, causing a gradual decline of the emission.
The UV magnitude requires $\sim 50-100$~Myr to fade by $\sim 3$ mag, thus remaining sensitive to star formation over relatively long timescales.

Hydrogen emission line EW ($\rm H\alpha$ and $\rm H\beta$, second panel), defined as the ratio between line luminosity and the underlying continuum emission, are powerful tracers of recent star formation activity. Hydrogen recombination lines are produced in HII regions ionized by short-lived massive OB stars, and trace the SFR over the past $\lesssim5$~Myr, as the ionizing photon output drops rapidly once these stars evolve off the main sequence.
In contrast, the optical continuum is primarily sensitive to the accumulated stellar mass and longer-lived stellar populations. As a result, hydrogen line EWs broadly trace the specific SFR. Following a starburst, the EW remains high for the first $\sim5$~Myr, before declining rapidly as the massive stars die. For example, the $\rm H\alpha$ EW decreases by a factor $\sim10$ between 3 and 10 Myr\footnote{We note that we do not include binary stars, which could prolong the Balmer-line emission beyond $\sim 10$~Myr \citep{Stanway2016}. However, at low metallicity binaries emit only $\lesssim$20\% of their ionizing photons after 10 Myr \citep{Ma2016} and on timescales still shorter than those of our SFH fluctuations ($\sim50$~Myr, Sec.~\ref{Sec:fiducial}), we therefore do not expect this slight delay to significantly affect our conclusions.}.

The UV continuum slope (third panel) is defined as $\beta_{\rm UV}$, where $f_\lambda \propto \lambda^{\beta_{\rm UV}}$. 
We measure $\beta_{\rm UV}$ by fitting the slope of the rest-frame continuum between 1600-2500~\angstrom, following the standard definition \citep[e.g.][]{Calzetti1994}.
The UV slope is sensitive to the presence of young stellar populations, nebular continuum emission, and dust attenuation, which can redden the intrinsic spectrum. Here we show the evolution of the intrinsic $\beta_{\rm UV}$, i.e. without any dust attenuation. Following a starburst, galaxies exhibit very blue slopes ($\beta_{\rm UV} \lesssim -2.5$) during the first $\sim10$~Myr, when the emission is dominated by young massive stars and nebular continuum emission. Around $\sim5-10$~Myr, the nebular contribution, which makes the UV slope less steep than the stellar continuum, rapidly declines. As a result the overall UV slope becomes temporarily even bluer.
At later times, as the stellar population ages and the short-lived massive O/B stars die, the UV continuum becomes dominated by longer-lived lower mass stars and the spectrum progressively reddens, reaching $\beta_{\rm UV} \sim -2$ after $\sim100$~Myr.

The fourth panel shows the Balmer break, defined as the flux ratio $F_\nu(4200\angstrom)/F_\nu(3500\angstrom)$ \citep[e.g.][]{Binggeli2019, Vikaeus2024}. Following a starburst, the Balmer break first exhibits a bump around $\sim5$~Myr due to strong nebular continuum emission powered by massive stars. After $\sim10$~Myr, it increases steadily as the stellar population ages, exceeding unity after $\sim20$--30~Myr when A-type stars begin to dominate the spectrum. The Balmer break is therefore primarily driven by the relative contribution of older stellar populations compared to young, massive stars: a weak break indicates a spectrum dominated by recent star formation, while a stronger break reflects an increasing contribution from evolved stars.

\begin{figure*}[tp]
    \centering
    \includegraphics[width=\linewidth]{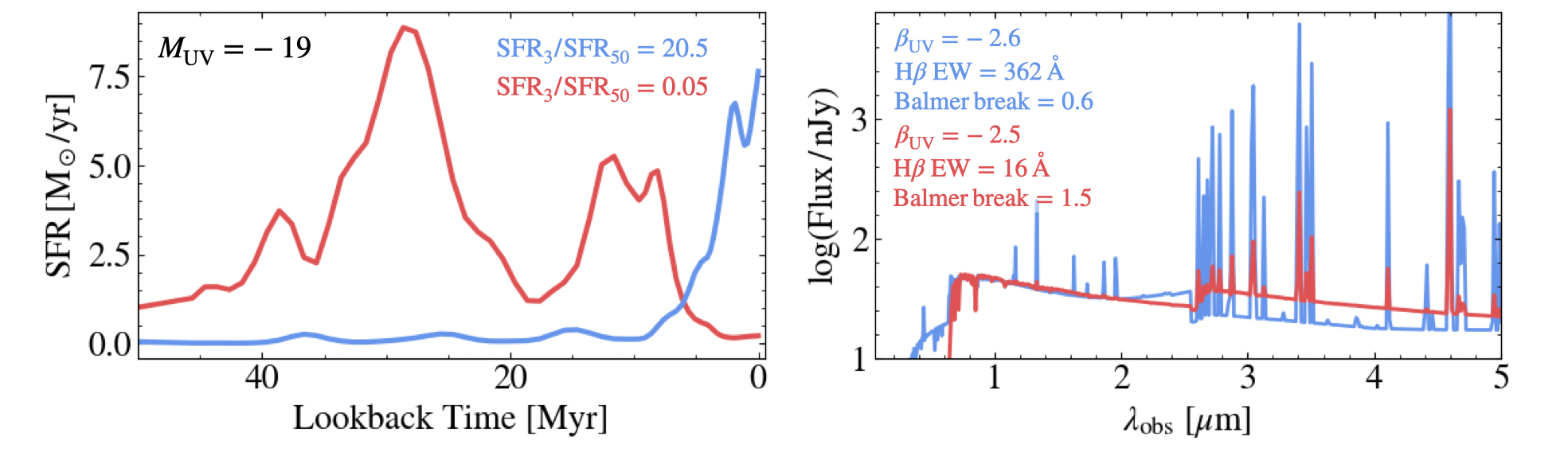}
    \caption{Star formation histories (left) and spectral energy distributions (right) of two example bursty galaxies at $z=6$. The two galaxies have the same UV magnitude $M_{\rm UV} = -19$, but markedly different histories: one is observed during an upturn phase (blue), while the other is in a downturn phase (red). Despite having identical UV magnitudes, their spectra differ significantly: the burst galaxy shows a steeper UV slope, a weaker Balmer break, and a stronger $\rm H\beta$ equivalent width compared to the downturn system.
    }
    \label{fig:sfh_examples}
\end{figure*}

While these trends provide useful intuition, bursty star formation histories complicate the interpretation of galaxy observables, since different tracers retain memory of star formation over distinct and overlapping timescales.
To illustrate this, we present two representative examples of bursty SFH and their emission in Fig.~\ref{fig:sfh_examples} for two galaxies with same $M_{\rm UV}= -19$. The SFHs of two $z=6$ galaxies are constructed using the method described above, with parameters that will be calibrated in Sec.~\ref{Sec:fiducial}. The left panel shows their star formation histories over the last 60~Myr, highlighting that systems with identical UV luminosities can exhibit very different SFHs. The galaxy in red is in a declining phase, with its SFR decreasing from $\sim5~\msunyr$ to nearly zero over the past $\sim10$~Myr, while the blue galaxy is experiencing a recent burst of star formation.
A commonly used way to characterize this burstiness is to quantify the ratio between the star formation rate averaged over the last 3~Myr and that averaged over the last 50~Myr, $\rm SFR_3/SFR_{50}$ \citep[see e.g.][]{Cole23, Looser23b, Endsley25, Kokorev2025, Tang2025, Heintz2026, Stoffers26}. Values above unity indicate a recent increase in star formation relative to the longer-term average, while values below unity trace a downturn or quenching phase. In the two examples shown here, this ratio has values of $0.05$ for the declining system and $20.5$ for the bursting one.

The diversity of SFHs at fixed $M_{\rm UV}$ propagates directly into differences in spectral properties. The right panel of Fig.~\ref{fig:sfh_examples} shows the corresponding spectra, highlighting variations in Balmer break, Balmer line EW and UV slope driven by the different recent star formation activity. The bursting galaxy shows stronger emission-line equivalent widths, a Balmer jump, and slightly bluer UV slope, all reflecting the dominance of young stellar populations. Conversely, the galaxy in a declining phase shows weaker line emission, a well developed Balmer break and slightly redder UV slope. Studying the distributions of these observables as a function of $M_{\rm UV}$ across large samples can therefore be key to constraining burstiness in high-redshift galaxies.

\subsection{Generating galaxy populations}

Having described how we model the evolution and emission of individual halos, the next step is to simulate a population of galaxies to predict global properties, scaling relations, and the distributions of observables across different redshifts.

We assume a standard $\Lambda$CDM cosmology with dark-matter halos distributed according to the halo mass function (HMF), for which we adopt the form of \cite{Reed07}\footnote{We verified that adopting different halo mass functions such as those from \cite{Yung2024_gureft} produces negligible differences in our results and does not affect the conclusions of this work.}.
At each redshift we draw $\gtrsim10^7$ halos from the HMF.
For each halo, we generate a random SFH with the PSD method described above, and forward model the corresponding SED.

We impose a lower limit on the halo masses capable of forming stars, set by the atomic cooling threshold corresponding to virial temperatures of $T_{\rm vir} \sim 10^4$~K \citep{BrommYoshida11}. We note that it is crucial to simulate halos down to this limit to be mass complete at fixed UV luminosity\footnote{While molecular cooling can in principle enable star formation in lower-mass minihalos, these systems are expected to be strongly suppressed by Lyman Werner background and to provide negligible contribution to the UV-bright high-$z$ galaxy population \citep[e.g.][]{Wise2012}}. In fact, due to stochasticity in their SFH, low-mass halos may experience upward fluctuations in star formation and contribute to the galaxy population at high luminosities.

\begin{figure*}[tp]
    \centering
    \includegraphics[width=\linewidth]{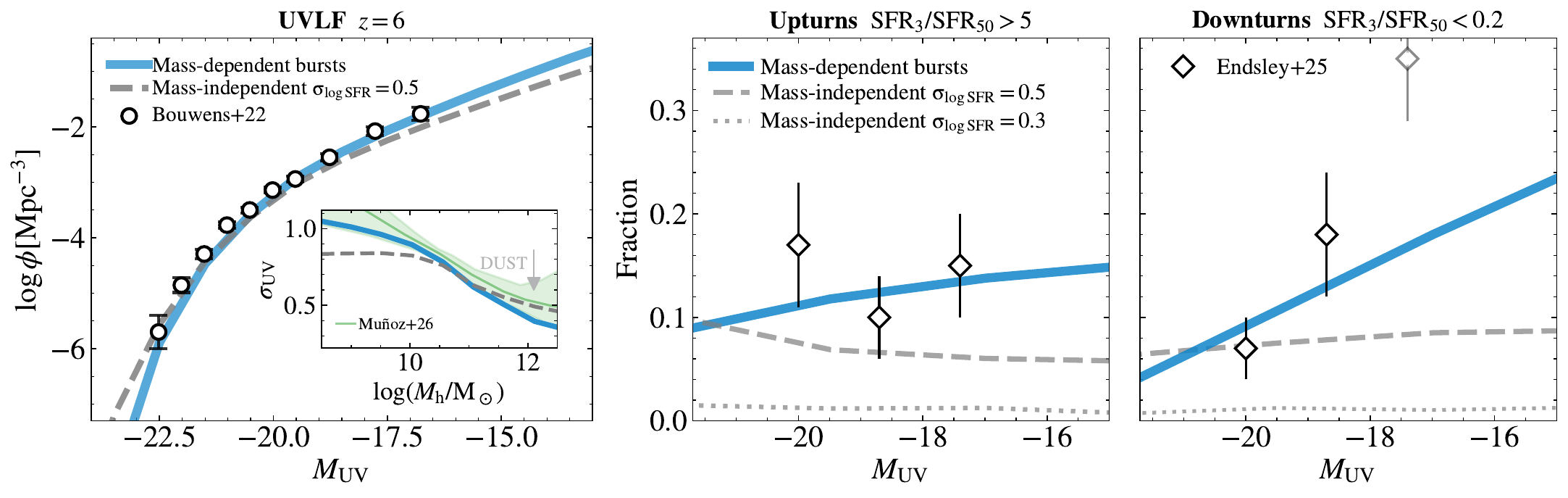}
    \caption{
    Results and prediction for the fiducial mass-dependent model at $z=6$, and comparison with a model without mass-dependent scatter.
    \textit{\textbf{Left:}} UV luminosity function for the mass-dependent burstiness model (blue) and a mass-independent model with $\sigma_{\log\rm SFR}=0.5$ (dashed grey), compared to \citet{Bouwens22lf}. \textit{\textbf{Inset:}} resulting UV scatter $\sigma_{\rm UV}(M_h)$, the mass-dependent model gives $\sigma_{\rm UV}(10^{10}\msun)\sim0.9$~mag and declines toward higher $M_h$ due to dust attenuation, compared with the results of \cite{Munoz2026}.
    \textbf{\textit{Right:}} fractions of galaxies in upturns (SFR$_3$/SFR$_{50}>5$) and downturns ($<0.2$) as a function of $M_{\rm UV}$, compared with data from \cite{Endsley25}. Only the mass-dependent model reproduces the observed rise in downturn fraction toward faint $M_{\rm UV}$, showing that mass-dependent stochasticity is required to match the data.
    }
    \label{fig:z6model}
\end{figure*}
\section{Bursty star formation at \lowercase{$z=6$}}\label{Sec:fiducial}
To investigate how bursty star formation shapes galaxy properties across cosmic time, we here first establish a fiducial model calibrated using the well-constrained galaxy population at $z=6$ (Section~\ref{subsec:z6_calibration}). We then test its predictions against multiple observational diagnostics at the same redshift (Section~\ref{subsec:z6_tests}).

\subsection{Model calibration at $z=6$}\label{subsec:z6_calibration}
We aim to identify a set of parameters that best reproduce the observed UV luminosity function and the inferred stochasticity of star formation histories at $z=6$. For this purpose, we use the UV luminosity function from \citet{Bouwens22lf} together with estimates of the SFR$_3$/SFR$_{50}$ ratio distributions from \citet{Endsley25}, which provide direct constraints on star formation variability in $z=6$ galaxies. 

\citet{Endsley25} analyzed a sample of 368 galaxies at $z\sim6$ across a wide magnitude range $(-22\lesssim M_{\rm UV}\lesssim-16)$, using JWST/NIRCam photometry to infer recent SFH from $\rm H\alpha$-to-UV luminosity ratios combined with SED fitting. They find that fainter galaxies have broader distributions of $\rm H\alpha$-to-UV, hinting at increased burstiness in lower mass halos. We note that the nature of the population of weak emission line sources is not fully established: low nebular emission line EW can arise not only from phases of suppressed star formation but also from other factors, such as high ionizing photon escape fractions or strong differential dust extinction \citep[e.g. see][]{Endsley25}. 
However, SFH are likely to play a dominant role in suppressing EW \citep[e.g.,][]{Shivaei2018,Endsley2023,Chen2026,Harikane26}.
Throughout this paper, we thus assume that the observed emission-line EWs, and the inferred SFR ratios, primarily trace the SFH. We discuss this assumption further in Sec.~\ref{sec:discussion}.

We explore a grid of models varying the burstiness and star formation efficiency parameters introduced in Section~\ref{Sec:methods}, in the ranges $\sigma_{10} \in [0.3, 1.5]$, $\tau \in [0.1, 100]$~Myr, and $\epsilon_{\star,0} \in [0.01, 1]$. We find a fiducial model with $\sigma_{10}=0.6$, $\tau=50$~Myr, and $\epsilon_{\star,0}=0.03$ provides the best match to the $z\sim6$ observational constraints.
Notably, the selected fluctuation timescale is consistent with values found in simulations, where SFHs are modulated by stellar feedback processes such as supernovae \citep[e.g.][]{Ma18, Gelli20, Gelli23, Pallottini23}. The values are also consistent with recent results by \cite{Munoz2026}, who used a similar parametrization of the time-variable SFH.

We show the fiducial mass-dependent model resulting from the calibration in Fig.~\ref{fig:z6model}. The left panel shows the UVLF compared with data from \cite{Bouwens22lf}.
To assess the impact of the assumed mass-dependent scatter, we compare the results from the fiducial mass-dependent model with a model with no mass dependence, with $\sigma_{\rm \log SFR} = 0.5$, shown with dashed curve.
Both models reproduce the observed UVLFs reasonably well. Their shapes differ slightly due to the different halo mass dependence of burstiness: in the mass-dependent model, the larger scatter in lower-mass halos enhances the faint end of the UVLF more than the bright end, resulting in a slightly steeper luminosity function.

\begin{figure}
    \centering
    \includegraphics[width=0.73\linewidth]{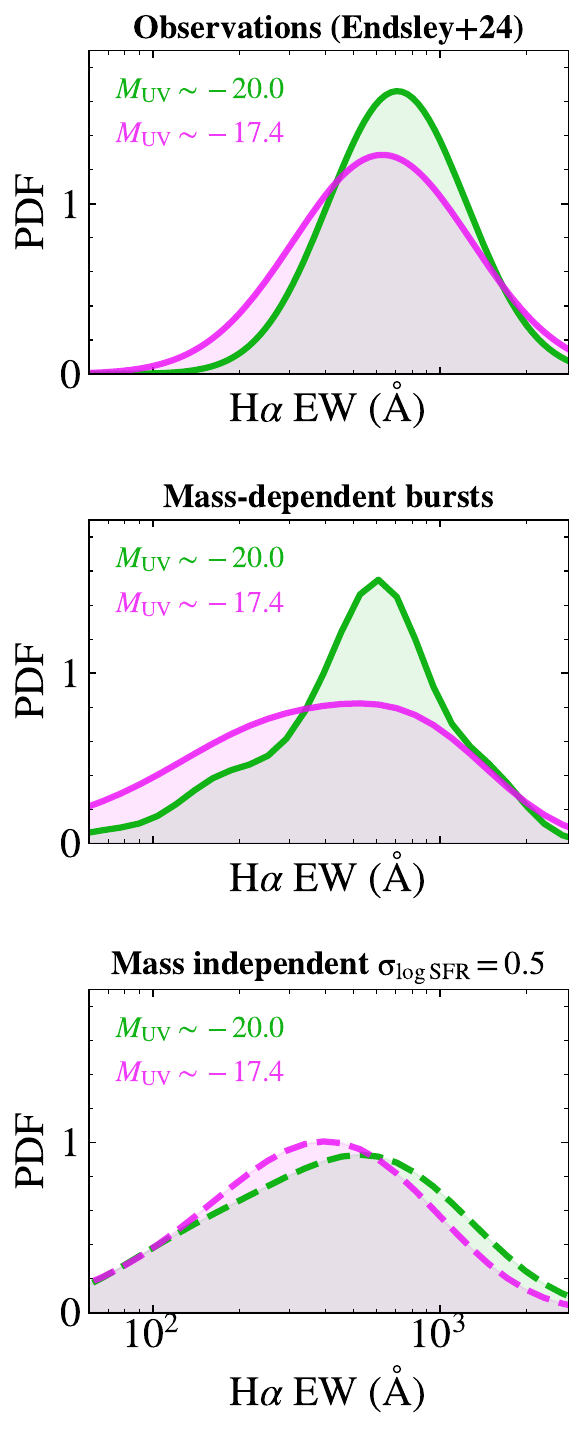}
    \caption{
    Probability distribution functions of H$\alpha$ EW in two magnitude bins ($M_{\rm UV}=-20$ and $-17.4$), for observations \citep[top,][]{Endsley23b}, the mass-dependent model (middle), and the mass-independent model (bottom). The observations show both a shift to lower median EW and a broadening of the distribution from the bright to the faint bin; the mass-dependent model reproduces both trends, while the mass-independent model does not reproduce the broadening.
    \label{fig:z6_HaEW}
    }
\end{figure}

The inset shows the resulting UV scatter $\sigma_{\rm UV}$ as a function of halo mass, defined as the standard deviation of the $M_{\rm UV}$ distribution at fixed $M_h$ and computed in 0.5 dex mass bins. The UV scatter reflects, but is not identical to, the assumed mass-dependent scatter in $\log \mathrm{SFR}$ (Equation~\ref{eq:mass_dep}), since the UV luminosity is sensitive to star formation variations over longer timescales. We find $\sigma_{\rm UV}(10^{10} M_\odot) \sim 0.9$\,mag for our mass-dependent model,\footnote{We note this value is slightly lower than assumed in \citet{Gelli24b}, where a Gaussian distribution of $P(M_{\rm UV}|M_h)$ is adopted. This difference arises because, in the present model, the mapping between $M_h$ and $M_{\rm UV}$ is not imposed a priori but emerges self-consistently from stochastic star formation histories, resulting in a non-Gaussian and skewed $P(M_{\rm UV}|M_h)$ distribution. We discuss this distribution shape in more detail in  Appendix~\ref{Appendix:pdf_Muv}.}  consistent with results from \cite{Munoz2026}.
The mass-independent model instead yields an approximately flat relation with $\sigma_{\rm UV}\sim0.8\rm\,mag$ for $M_h\lesssim10^{11}\rm M_\odot$. At high halo masses, we see a decline of $\sigma_{\rm UV}$ driven by dust attenuation, which preferentially suppresses UV-bright galaxies during starburst phases, reducing the amplitude of upward luminosity fluctuations and hence the UV scatter. In the absence of dust attenuation, for the mass independent model the $\sigma_{\rm UV}-M_h$ relation remains flat over the full halo mass range.

The two panels on the right show the fractions of galaxies in recent upturns (left) and downturns (right) of star formation, defined as those with ratios of SFR$_3$/SFR$_{50}>5$ and $<0.2$ respectively, following the definition by \cite{Endsley25} to compare with their data, as a function of $M_{\rm UV}$.
In the mass-dependent model, we find an upturn fraction of $\sim 10-15\%$, with a mild increase toward fainter UV magnitudes, while the downturn fraction increases more strongly toward the faint end. These trends with $M_{\rm UV}$ are driven by mass-dependent burstiness: lower-mass halos, which on average host fainter galaxies, exhibit stronger stochasticity and are therefore more likely to experience both enhanced and suppressed periods of star formation.

We find our mass-dependent model reproduces the general trends inferred by \cite{Endsley25}, with very good agreement for the fraction of upturning systems across the full magnitude range and for downturning systems down to $M_{\rm UV} > -18$. In the faintest magnitude bin ($M_{\rm UV} \sim -17$, shown in lighter shade) our model underproduces downturn systems relative to the fraction inferred by \citet{Endsley25}. With our current prescriptions, we find no parameter combination can simultaneously reproduce this faint bin together with the upturn and downturn fractions at brighter luminosities and the UVLF shape. 
This may indicate that, particularly in faint galaxies, which are more likely to reside in the lowest mass halos, star formation may undergo longer suppression phases or more rapid post-burst quenching than captured by our single-timescale stochastic model \citep[e.g.][]{Gelli23}.
We also note, however, that a high ionizing photon escape fraction can also reduce H$\alpha$ EW (used to infer short timescales SFR), and may therefore contribute to an enhanced inferred fraction of galaxies in downturn phases \citep{Endsley23b}.
We discuss this further in Section~\ref{sec:discussion}.

By contrast, the mass-independent model (shown in gray) does not describe the observations well. This model yields upturn and downturn fractions that are independent of UV magnitude, and fails to reproduce the increase in downturn fraction toward fainter galaxies.
We also show, for comparison, a model with a lower level of stochasticity, $\sigma_{\log \rm SFR}=0.3$ (dotted gray curve), comparable to the level of scatter expected from stochastic halo accretion alone \citep[e.g.][]{Ren18}. 
This model predicts upturn and downturn fractions below $\lesssim5\%$ across all magnitudes, substantially underestimating the observed values. This demonstrates that the stochasticity associated with halo accretion alone is insufficient to reproduce the observed data. A substantial level of burstiness in star formation histories and its mass dependence are required to reproduce the observed fractions of galaxies in recent upturns and downturns.

\subsection{Testing $z=6$ spectral observables predictions}\label{subsec:z6_tests}

To further validate the $z=6$ model, we compare its predictions against two observables which were not used in the calibration.
We compare our predictions for the distribution of H$\alpha$ equivalent widths (EW, Section~\ref{subsec:z6_tests_Ha}) and intrinsic UV slope, $\beta_\mathrm{UV}$ (Section~\ref{subsec:z6_tests_beta}), both as a function of UV magnitude, to recent JWST observations.
In each case, we expect the predictions of the models as a function of UV magnitude to be most important for understanding the impact of mass-dependent stochasticity.

\subsubsection{H$\alpha$ equivalent width}\label{subsec:z6_tests_Ha}

Fig.~\ref{fig:z6_HaEW} shows the distributions of H$\alpha$ equivalent widths in two UV magnitude bins ($M_{\rm UV} = -20$ to $-17.4$). The top panel shows the observed H$\alpha$~EW distributions as derived by \cite{Endsley23b} from deep NIRCam imaging of more than 250 galaxies from JADES. The middle and bottom panels show the distributions as predicted by our fiducial mass-dependent model and by the mass-independent model respectively.
The observations (top) show two trends in the H$\alpha$~EW distributions from the UV-bright to faint bin: a slight shift toward lower EW and a broadening of the distribution. 

Our mass-dependent model (middle) successfully reproduces both trends.
The shift of the median to lower EW reflects the lower SFR of less massive halos (due to lower accretion rates), which on average host fainter galaxies.
The broadening of EW distributions for fainter galaxies arises naturally in the presence of mass-dependent burstiness: lower-mass halos, hosting on average fainter galaxies, exhibit stronger stochasticity and star formation rates variations, hence giving rise to broader distributions of H$\alpha$~EW.
In contrast, the mass-independent model (bottom panel) reproduces a shift to lower EW at fainter UV-magnitudes, but does not reproduce the broadening of EW distribution. Because the level of burstiness is the same for all halo masses in this model, the shape of the EW PDF remains approximately unchanged between the two magnitude bins. Any small differences in shape between the two $M_{\rm UV}$ bins are consistent with statistical fluctuations due to finite sample size.

\subsubsection{UV slopes}\label{subsec:z6_tests_beta}

\begin{figure}
    \centering
    \includegraphics[width=\linewidth]{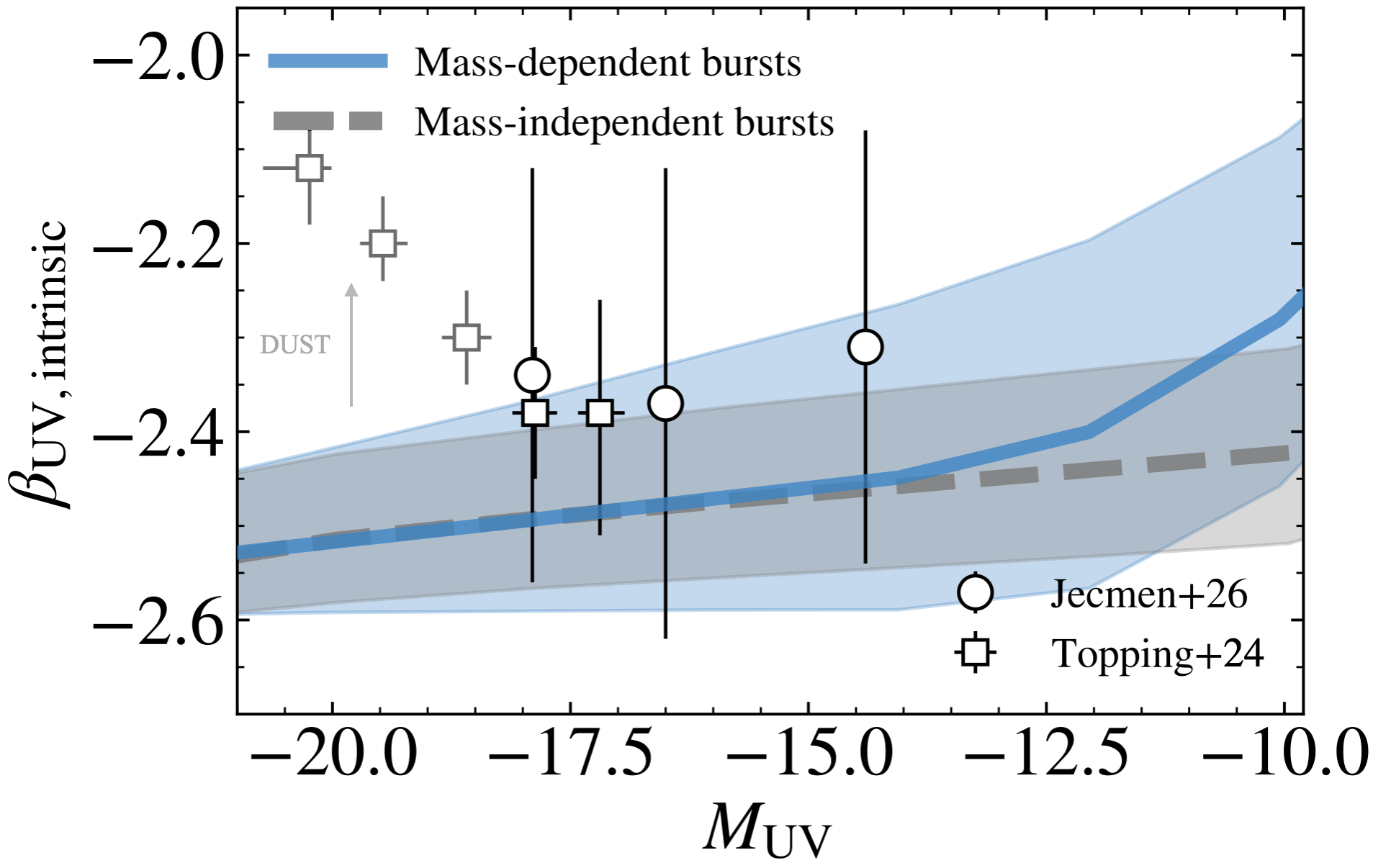}
    \caption{Intrinsic (no-dust) UV slopes $\beta_{\rm UV}$ as a function of UV magnitude at $z=6$, as predicted from the mass-dependent model (blue) and mass-independent burstiness model (dashed grey). JWST observations from \cite{Topping2024} and \cite{Jecmen2026} are also shown. 
    The systematic offset between the models and the observed $\beta_{\rm UV}$ values is mostly due to dust attenuation, which is not included for the models in the plot. Both mass-dependent and -independent models predict a near-flat relation at $M_{\rm UV}\lesssim-15$, but only the mass-dependent model predicts progressive reddening and increasing scatter toward fainter magnitudes ($M_{\rm UV}\gtrsim-15$), qualitatively consistent with the observed turnover.
    }
    \label{fig:z6_betaUV_Muv}
\end{figure}

In Fig.~\ref{fig:z6_betaUV_Muv} we explore the predicted intrinsic UV slopes, i.e. prior to dust attenuation, as a function of UV magnitude. By excluding dust, we isolate the contribution of star formation variability to the UV slope. We also compare to observations from JWST data from \cite{Topping2024} and \cite{Jecmen2026}.

Both the mass-dependent (blue solid line) and mass-independent models (dashed gray line) predict an almost flat relation for $M_{\rm UV} \lesssim -15$, with brighter galaxies having slightly bluer intrinsic UV slopes due to their higher average accretion and star formation rates.
Compared to recent JWST observations at $z\sim6$ \citealt{Topping2024,Jecmen2026}, our models both produce bluer median UV slopes than observed for galaxies brighter than $M_{\rm UV} < -18$.
At the bright-end ($M_{\rm UV} < -18$) the observed behavior is driven by stronger dust attenuation in more luminous galaxies. In the absence of dust attenuation, our model predicts an intrinsically nearly flat $\beta_{\rm UV}$–$M_{\rm UV}$ relation over this magnitude range.

At the faint end ($M_{\rm UV} \gtrsim -15$), the mass-dependent model predicts a progressive reddening, with $\beta_{\rm UV}$ increasing more rapidly toward lower luminosities.
This behavior arises naturally from mass-dependent burstiness: faint galaxies reside in more bursty low-mass halos, where an increasing contribution from downturn phases shifts the median toward redder intrinsic UV slopes. 
Such faint galaxies have long been out of reach observationally.
Excitingly however, recent JWST observations of strongly lensed galaxies from the GLIMPSE program show evidence for a similar faint-end reddening, with a turnover at $M_{\rm UV} \gtrsim -17$ \citep{Jecmen2026}. A flattening between $M_{\rm UV} \sim -18$ and $-17$ is also reported from deep JADES observations by \citet{Topping2024}; and \citet{Felicioni2026} find a consistent trend across multiple lensed fields extending to $M_{\rm UV} \sim -19$. Although the observed $\beta_{\rm UV}$ values are systematically redder than our intrinsic predictions, as expected due to dust attenuation, the detection of an enhanced reddening toward faint magnitudes is qualitatively consistent with mass-dependent stochastic star formation histories.

In the mass-dependent model, the intrinsic $\beta_{\rm UV}$ distribution shows a significant dispersion around the median (shaded region shows 16–84 percentile range). The scatter increases toward fainter magnitudes, as a consequence of mass-dependent burstiness: lower-mass halos hosting fainter galaxies are more strongly stochastic and therefore exhibit a wider range of UV slopes.
By contrast, the mass-independent model shows approximately constant scatter and no systematic broadening toward faint magnitudes, since the burst amplitude does not depend on halo mass.
A large scatter in the $\beta_{\rm UV}$–$M_{\rm UV}$ relation is measured in $z\sim6$ observations \citep[e.g.][]{Roberts-Borsani2024, Saxena2024, Topping2024}, with $\beta_{\rm UV}$ values spanning $-2.8 \lesssim \beta_{\rm UV} \lesssim -1.5$. 
This range is much broader than our model predictions: likely due to additional sources of variations such as dust attenuation and high escape fractions. However, our model suggests that burstiness alone contributes for part of the observed scatter, producing variations of $\beta_{\rm UV}$ of about $\sim 0.2$ at fixed $M_{\rm UV}$.

Having established that the mass-dependent model successfully reproduces many statistical properties of galaxies at $z\sim6$, we now test whether the same model naturally reproduces the observed evolution of burstiness toward $z=10$.

\begin{figure*}[tp]
    \centering
    \includegraphics[width=0.9\linewidth]{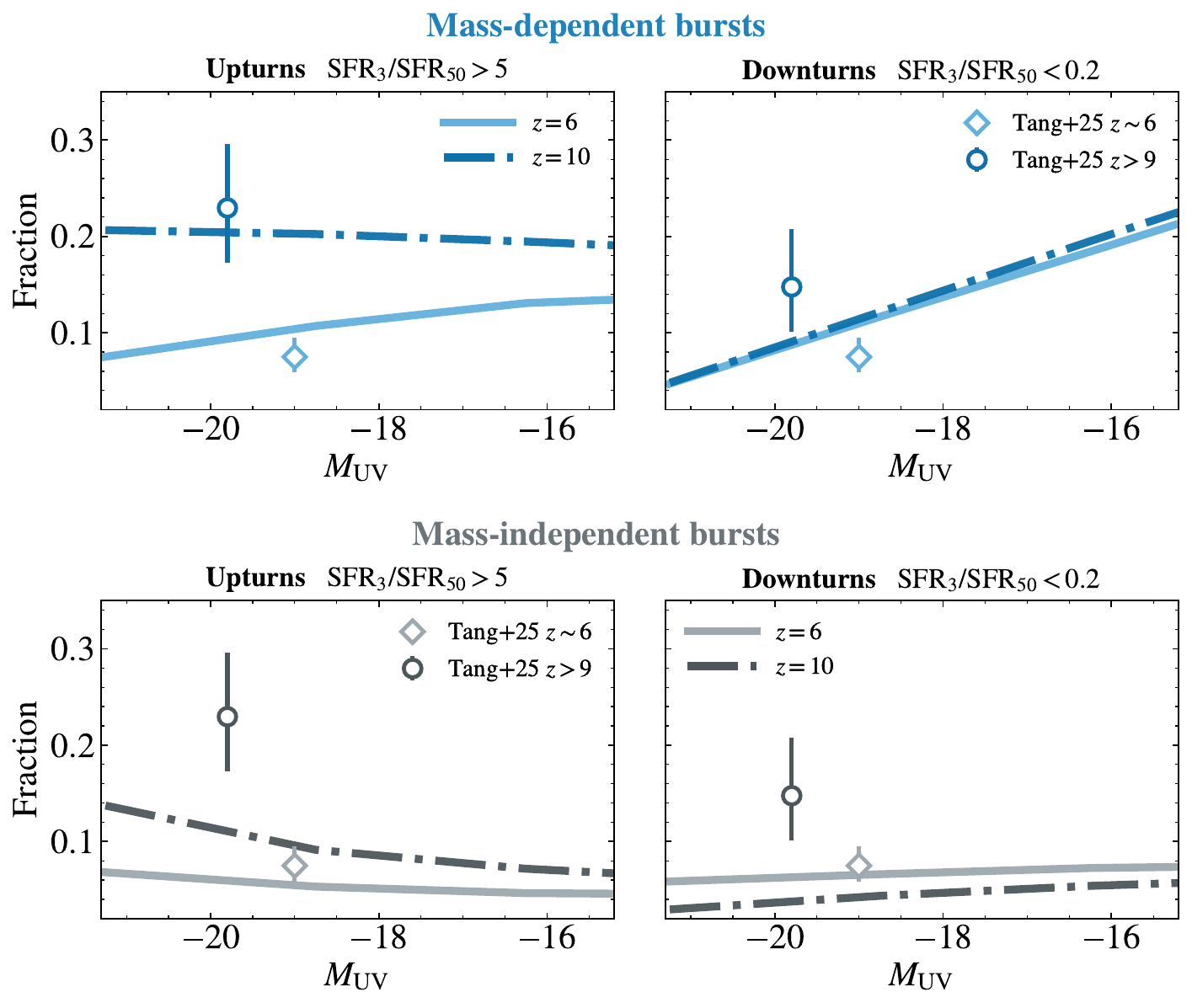}
    \caption{Redshift evolution of upturn and downturn fractions from $z=6$ to $z=10$. Fractions of galaxies in star formation upturns (SFR$_3$/SFR$_{50}>5$, left) and downturns ($<0.2$, right) as a function of $M_{\rm UV}$, for the mass-dependent (top) and mass-independent (bottom) models, compared to \citet{Tang2025}. The mass-dependent model reproduces the observed rise in upturn fractions from $\sim10\%$ at $z\sim6$ to $\sim25\%$ at $z\sim10$; while the mass-independent model predicts a much weaker increase and fails to match the data.
    }
    \label{fig:sfr_ratios_evol6_10}
\end{figure*}

\section{Redshift \lowercase{$z=6$} to \lowercase{$z=10$} evolution}\label{sec:z6_10_evol}

We now apply the $z=6$-calibrated model parameters at $z=10$ to explore the predicted redshift evolution and test whether the model reproduces JWST burstiness measurements at these epochs.
We once again focus on the fraction of upturns ($\rm SFR_3/SFR_{50}>5$) and downturns ($\rm SFR_3/SFR_{50}<0.2$) as a function of $M_{\rm UV}$, here focusing on their evolution from 6 to 10.

Fig.~\ref{fig:sfr_ratios_evol6_10} shows the prediction for the fiducial model (top row) and for the mass-independent model (bottom row). 
As increasingly large JWST NIRSpec prism spectroscopy samples are now making it possible to infer star formation histories and estimate these fractions at high redshifts, we can compare with recent observational results. In particular, we compare our predictions with \cite{Tang2025}, who analyzed a sample of 61 $-22<M_{\rm UV}<-17$ galaxies with prism spectroscopy across $z=9-14$ and compared their properties against a sample of 401 galaxies at $6<z<9$.

The fiducial model at $z=6$ shows a fraction of upturns (top left) between $\sim 8-14\%$ across the magnitude range $-21<M_{\rm UV}<-15$, increasing to $\sim 20\%$ at $z=10$.
This redshift evolution is primarily driven by mass dependent burstiness: at fixed $M_{\rm UV}$, galaxies are hosted by lower-mass halos toward higher redshift, which are intrinsically more bursty and therefore more likely to exhibit high $\rm SFR_3/SFR_{50}$. The increase in accretion rates at earlier times further contributes to this trend, driving more rapidly rising star formation histories at higher-z and therefore also higher average $\rm SFR_3/SFR_{50}$ ratios. This latter effect is stronger for brighter galaxies, which are hosted by more massive halos with higher accretion rates at fixed redshift, resulting in a flatter dependence of the upturn fraction with $M_{\rm UV}$.

Comparing with observations, \cite{Tang2025} inferred star formation histories by fitting the observed spectra and used these to estimate the $\rm SFR_3/SFR_{50}$ ratios, adopting the same definition as in our model. They find that the fraction of upturns increases from $\sim10\%$ to $\sim25\%$ from $z\sim6$ to $z\sim10$, in agreement with our model's prediction. This agreement supports our picture in which galaxies at higher redshift are progressively burstier, as they are hosted by lower-mass, more stochastic halos, resulting in higher upturn fractions, an effect further amplified by the rising accretion rates at high-$z$.

In contrast, the mass independent model (bottom) predicts a much weaker increase in the upturn fraction from $z=6$ to $z=10$. In this model, the increase is driven purely by the rise in accretion rates, pushing the average $\rm SFR_3/SFR_{50}$ ratios to higher values.
However, this increase is not sufficient to match the observed upturn fraction at $z\sim10$.

The panels on the right show the fraction of galaxies in downturn phases, those that have experienced a recent suppression in star formation and therefore undergo periods of reduced UV luminosity.
In the fiducial mass-dependent model (top right), the downturn fraction increases toward fainter luminosities, reflecting the mass dependence of burstiness. 
However, in contrast to upturn fractions, the downturn fraction shows little evolution from $z=6$ to $z=10$.
This is because, at fixed $M_{\rm UV}$, galaxies in downturn phases are preferentially systems that have been scattered down in UV luminosity from more massive halos.
At higher redshift, a fixed UV luminosity corresponds to lower-mass and more bursty halos, increasing the probability of suppression episodes.
At the same time, higher accretion rates at higher redshift lead to higher average $\rm SFR_3/SFR_{50}$ ratios, reducing the probability of having low $\rm SFR_3/SFR_{50}$ systems.
These two effects approximately compensate, leading to a nearly redshift-independent downturn fraction evolution.
The observed downturn fractions inferred by \citet{Tang2025} are broadly consistent with our predicted trends, though they may hint at a mild increase in downturns at earlier epochs compared to our model.

Conversely, the mass-independent model (bottom right) does not reproduce the data. The downturn fraction is constant with $M_{\rm UV}$, and it decreases from $z=6$ to 10 by a factor $\sim 1.5$. This is because, without mass-dependence, the $\rm SFR_3/SFR_{50}$ distribution shifts to higher values from $z=6$ to 10, while retaining the same scatter, hence reducing the fraction of galaxies below the $\rm SFR_3/SFR_{50}<0.2$ threshold.
Importantly, this shows the downturn fractions are mainly determined by the amplitude and mass-dependence of stochasticity, rather than accretion rates.
As we will discuss in Section~\ref{sec:discussion}, this makes the downturn fraction a particularly sensitive tracer of bursty star formation.

Overall, the results presented in this and the previous section demonstrate that: (i) a high level of burstiness is required to reproduce the observed fractions of upturns and downturns and the distributions of H$\alpha$ EW, and (ii) a mass-dependent burstiness is necessary to reproduce the observed trends with $M_{\rm UV}$ and redshift.
Having validated the fiducial model with mass-dependent burstiness at $z=6-10$, we now explore the model's predictions for key observables at $z=14$.

\section{Star formation models at \lowercase{$z>10$}}\label{sec:highz}

The JWST discovery of luminous galaxies at $z=14$ poses a significant challenge to models of early galaxy formation that struggle to reproduce the high observed abundances. 
Motivated by the success of our model at $z\sim6-10$, here, we test whether the burstiness inferred from our $z=6$ calibration can account for the observed UVLF at $z=14$, and investigate possible modifications to the bursty star formation histories required to reconcile the predictions with the JWST observations.

\begin{figure*}[t!]
    \centering
    \includegraphics[width=\linewidth]{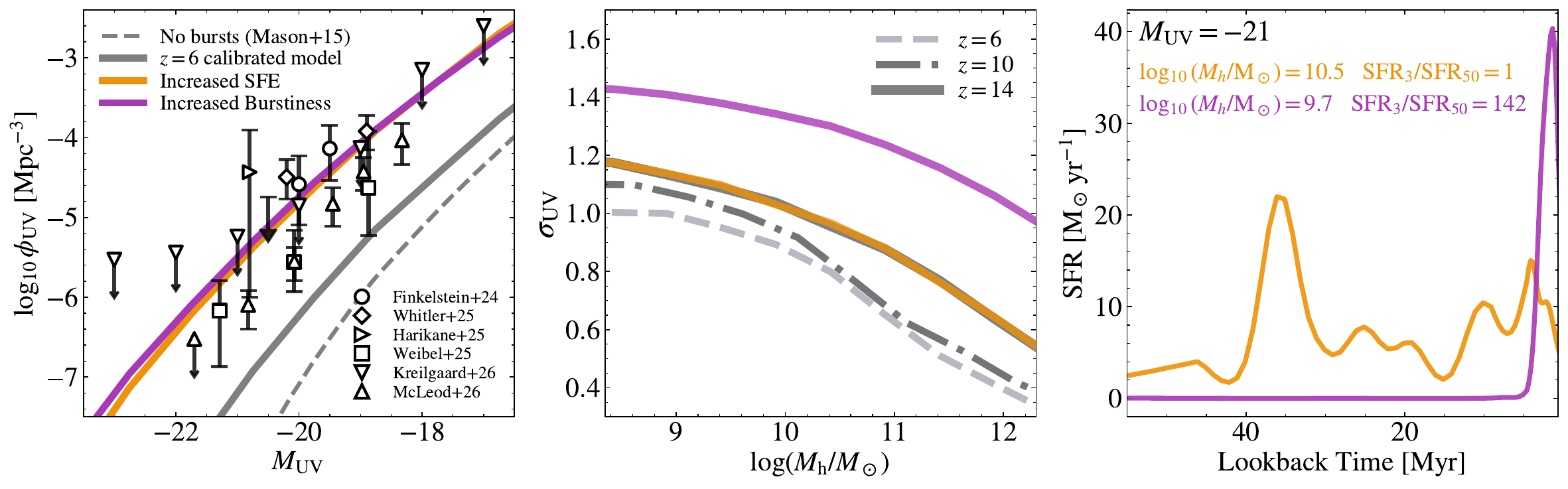}
    \caption{Different star formation models at redshift $z=14$. \textit{\textbf{Left:}} $z=14$ UVLF for the \textit{increased-burstiness} and \textit{increased-SF}E models, compared to JWST observations from \citep{Finkelstein23, Harikane2025, Weibel2025, Whitler25, Kreilgaard2026, McLeod2026}. The bursty model calibrated at $z=6$ (gray) enhances the number densities compared to the no-scatter case (dashed), but fails to reproduce the observed high number densities at $z=14$. \textit{\textbf{Center:}} UV scatter as a function of halo mass. The $z=14$ \textit{increased-burstiness} model shows the highest UV scatter, $\sigma_{\rm UV}(10^{10}\,M_\odot)\approx1.4$~mag, while \textit{increased-SFE} at $z=14$ retains the same scatter as the fiducial $z=6$-calibrated model, $\sigma_{\rm UV}(10^{10}\,M_\odot)\approx1$~mag. \textit{\textbf{Right:}} example SFHs for typical galaxies with the same UV magnitude $M_{\rm UV}=-21$ predicted by the two different models. The \textit{increased-burstiness} galaxy is caught in an ongoing strong burst in a lower-mass halo, while the \textit{increased-SFE} galaxy has had a more stable SFR throughout the past 50~Myr.
    }
    \label{fig:z14models}
\end{figure*}

Applying the $z=6$ calibrated fiducial model at higher redshifts, we find that it reproduces the observed UVLF remarkably well up to $z=11$. At $z=12-13$, the model continues to match the faint end but begins to underpredict the abundance of bright galaxies at $M_{\rm UV}\lesssim-19$ (Appendix~\ref{Appendix:z10UVLF}). By $z=14$, this discrepancy becomes substantial (solid gray curve in the left panel of Fig.~\ref{fig:z14models}). As also found in previous studies \citep[][]{Gelli24b, Munoz2026}, mass-dependent scatter enhances the abundance of UV-bright galaxies relative to models without stochasticity, such as the model of \cite{Mason15} (dashed line). This enhancement arises because lower-mass halos are more abundant relative to higher-mass halos at high redshift and also exhibit stronger burstiness, allowing them to contribute significantly to the bright end of the UVLF. However, the scatter calibrated at $z=6$ remains insufficient to reproduce the observed abundance of UV-bright galaxies at $z=14$ \citep[e.g.][]{Finkelstein23, Harikane2025, Weibel2025, Whitler25, Kreilgaard2026, McLeod2026}.

We therefore explore different scenarios designed to reproduce the UVLF at $z=14$, inspired by solutions proposed in recent theoretical work. As discussed in the introduction, these can be broadly divided into two classes according to how they impact galaxies' star formation histories: (i) increasing the median SFR at fixed halo mass (e.g. via higher star formation efficiency as shown by e.g. \citealt{Dekel23, Somerville2025}), and (ii) increasing the scatter in luminosity at fixed halo mass through more bursty star formation \citep[e.g.][]{Mason2023a, Munoz23}\footnote{Another explanation proposed in literature by \cite{Feldmann2025} is a shallow mass-dependence of the star formation efficiency. In Appendix~\ref{Appendix:sfe} we show that, under the same definition of efficiency, our mass-dependent burstiness model naturally produces a shallower effective $\rm{SFR}/(f_b\dot{M}_h) - M_h$ relation, because the larger stochastic fluctuations in low-mass halos enhance their mean SFR. We therefore interpret this effect as an emergent consequence of bursty star formation rather than as an additional independent scenario.}.

Within our framework, we model these two scenarios by varying the parameters for the normalization of the star formation efficiency $\epsilon_{\star,0}$ and for the bursts amplitude $\sigma_0$. 
Starting from the fiducial $z=6$ calibrated model, we set each of the two parameters to reproduce the observed $z\sim14$ UVLF from JWST, exploring the two scenarios:

\begin{itemize}
    \item \textbf{\textit{Increased SFE}\footnote{Here the term ``SFE'' refers to the parameter $\epsilon_{\star,0}$ in Equation~\ref{eq:sfe} which sets the normalization of the median $\mathrm{SFR}$ at fixed halo mass. This differs from other common literature definitions where SFE is proportional to the mean $\langle \mathrm{SFR} \rangle$ \citep[e.g.][]{Feldmann2025, Somerville2025}. See Appendix~\ref{Appendix:sfe} for how our mass-dependent burstiness model affects the mean SFR.}
    :} the normalization of the efficiency $\epsilon_{\star,0}$ is increased by a factor $\times~5$, while the stochasticity is the same as in the fiducial model $\sigma_{\rm UV}(M_h=10^{10}\msun)\approx1\rm\,mag$;\\

    \item \textbf{\textit{Increased burstiness}:} we increase the amplitude of stochastic fluctuations to $\sigma_{\rm 10} = 0.95$, with $\sigma_{\rm UV}(M_h=10^{10}\msun)\approx1.4\rm\,mag$, while leaving the efficiency unchanged. 
    \\
\end{itemize}

These models are illustrated in Fig.~\ref{fig:z14models}.
On the left we show the predicted UVLF at $z=14$ for the \textit{increased-SFE} and \textit{increased-burstiness} cases in orange and purple, that are tuned to match JWST data from \cite{Finkelstein23, Harikane2025, Whitler25, Kreilgaard2026}. More recent constraints from \cite{Weibel2025} and \cite{McLeod2026}, based on wide areas and pure parallel surveys, are also shown and, although they present lower number densities than the other studies, they remain above the prediction of the fiducial model calibrated at $z=6$, suggesting that some modification to the star formation histories is still required to match the observed abundance of bright galaxies at $z\sim14$.

The middle panel shows the UV scatter as a function of the halo mass for the fiducial model at $z=6$, 10 and 14, and for the two $z\sim14$ models.
The fiducial model, shown in grey, predicts $\sigma_{\rm UV}$ increases for lower masses, going from $\sim 0.5$\,mag to $\sim 1.2$\,mag in $M_h=10^{12}\msun$ to $10^{8}\msun$ halos at $z=6-10$. We note the model also predicts the \textit{amplitude} of $\sigma_{\rm UV}$ increases slightly at fixed halo mass at higher redshifts. 
This weak redshift dependent evolution of the UV scatter arises because UV luminosity traces star formation over $\sim50$~Myr, a timescale over which halos grow in mass.
A halo at a given mass at $z\sim14$ grows more over the past 50 Myr than a halo of the same mass at $z\sim6$. As a result, its UV light includes contributions from earlier phases when the halo was less massive than the same final-mass halo at $z=6$, and therefore more bursty. This naturally leads to a higher UV scatter at higher-z, despite our prescription of redshift independent SFR burstiness at fixed halo mass.

The $\sigma_{\rm UV}$ for the $z=14$  \textit{increased-SFE} model is identical to that of the fiducial $z=14$ model, as the UV scatter is not affected by the increase of the star formation efficiency. In contrast, the \textit{increased-burstiness} model has similar mass dependence of the UV scatter, but a systematic shift to higher values, ranging from $\sim1.5$ at low masses to $\sim1$ at the high-mass end.
This shows that, despite the fact that the $z\sim14$ UV luminosity functions predicted by the \textit{increased-SFE} and \textit{increased-burstiness} models are essentially indistinguishable, the scatter in UV luminosity at fixed $M_h$ differs between the two models.

The difference between the two models is driven by the underlying star formation histories. In the right panel of Fig.~\ref{fig:z14models}, we show examples of two typical SFHs for galaxies with $M_{\rm UV} \sim -21$, i.e. selected to be representative of systems hosted by halos near the median halo mass at that UV luminosity in each model.

In the \textit{increased-burstiness} model, a typical $M_{\rm UV} \sim -21$ galaxy has a star formation history consisting of a very strong young burst in the last $\sim5$~Myr, with little to no star formation previous to that event. It is hosted in a relatively low mass halo with temporarily boosted UV luminosity driven by the high SFR at the time of observation. Conversely, in \textit{increased-SFE} model the galaxy has instead lower SFR at the time of observation, but it has been forming stars from earlier on, the emission from slightly older stellar populations ($\sim20$~Myr) still contributing to the observed UV light.

Given the similarity of the predicted UVLFs, our goal is to establish which observables can best distinguish between the models.
As shown in Section~\ref{Sec:fiducial}, the distributions of key observables across large samples can differ systematically between models, providing a powerful way to constrain the burstiness of star formation histories and the physical origin of the UVLF excess.
We therefore investigate the predicted evolution of distributions of key observables from $z=6$ to $z=14$, and as a function of $M_{\rm UV}$ in these models. In particular, we focus on: $\rm SFR_3/SFR_{50}$ ratios (Sec.~\ref{subsec:sfr_ratio}), nebular emission line equivalent widths (Sec.~\ref{subsec:ew}), Balmer breaks (Sec.~\ref{subsec:balmer_break}), UV slopes (Sec.~\ref{subsubsec:uv_slopes}) and $z=17$ UVLF (Sec.~\ref{subsec:z20uvlf}).

\subsection{SFR ratios}\label{subsec:sfr_ratio}

In Sec.~\ref{Sec:fiducial} we introduced the $\rm SFR_3/SFR_{50}$ ratio as a powerful diagnostic for studying the statistical properties of recent star formation histories of galaxies. We here analyze the predictions of the $\rm SFR_3/SFR_{50}$ ratio at $z=14$, comparing the \textit{increased-burstiness} and the \textit{increased-SFE} scenarios introduced in the previous section.

\begin{figure}
    \centering
    \includegraphics[width=\linewidth]{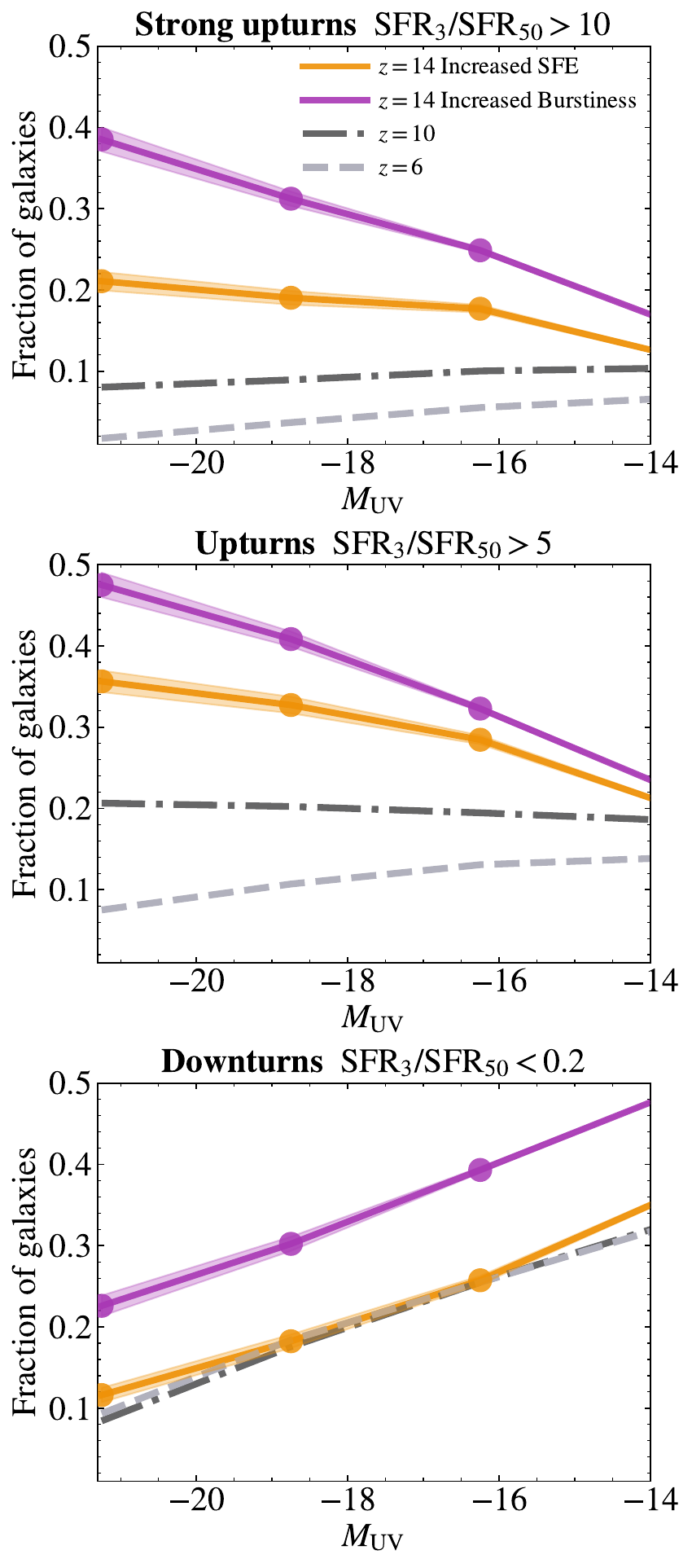}
    \caption{Fractions of galaxies in different phases of bursty star formation as a function of their UV magnitude. 
    The top panel shows the fraction of galaxies undergoing strong upturns, defined as $\rm SFR_3/SFR_{50} > 10$, the middle panel shows upturns with $\rm SFR_3/SFR_{50} > 5$, and the bottom panel shows galaxies in downturn phases, defined as $\rm SFR_3/SFR_{50} < 0.2$. For each phase, we show the expected redshift evolution, with gray lines indicate trends for the baseline model at $z=6$ and 10, while colored lines show the different $z=14$ star formation models.
    }
    \label{fig:sfr_ratios}
\end{figure}

We classify galaxies according to their $\rm SFR_3/SFR_{50}$ ratio as experiencing \textit{strong upturns} ($>10$), \textit{upturns} ($>5$) and \textit{downturns} ($< 0.2$).
Our aim here is to quantify how likely galaxies at different redshifts and luminosities are to be caught in phases of enhanced or suppressed star formation. To this end, each panel of Figure~\ref{fig:sfr_ratios} shows the fraction of galaxies in strong upturns, upturns and downturns respectively as a function of UV magnitude. When interpreting these trends, it is important to keep in mind that, at fixed halo mass, galaxies undergoing an upturn have boosted UV luminosities, while galaxies in downturns appear fainter than average.

As shown in Sec.~\ref{sec:z6_10_evol}, the fiducial model (here shown with grey curves) predicts a strong increase in the upturn fractions from $z=6$ to 10, consistent with recent JWST measurements \citep{Tang2025}, while the downturn fraction remains approximately constant over the same redshift range. We now examine how these trends continue toward $z=14$ in the \textit{increased-burstiness} and \textit{increased-SFE} scenarios, shown with the purple and orange curves respectively.
We note that the \textit{increased-SFE} scenario has the same level of stochasticity as the fiducial model (see middle panel of Fig.~\ref{fig:z14models}). Therefore, the evolution of the SFR ratios at $z=14$ is identical in the two models and, for this reason, we do not show the fiducial $z=14$ predictions in the following plots.

In both scenarios, the fraction of galaxies in strong-upturn (top) and upturn (middle) phases increases relative to $z=10$, with a stronger rise toward brighter magnitudes. This evolution reflects the combined effect of mass-dependent stochasticity and higher accretion rates at higher redshift. At fixed $M_{\rm UV}$, galaxies at higher-$z$ reside in lower-mass, more bursty halos, making upturns more likely. In addition, halo accretion rates increase more rapidly at earlier epochs, producing more rapidly rising star formation histories and therefore higher average $\rm SFR_3/SFR_{50}$ ratios, further increasing the fraction of galaxies with recent upturns.
This implies that, at fixed luminosity, galaxies are more likely to be observed in upturn phases at higher-$z$ than at lower-$z$. The effect becomes stronger at higher luminosities, where bursty phases become increasingly important.
Thus, the bright end of the UVLF is increasingly dominated by galaxies undergoing recent upturns in star formation at higher redshift.

Comparing the two models, the \textit{increased-burstiness} scenario predicts a significantly larger fraction of galaxies in upturn phases than the \textit{increased-SFE} model. In particular, $\sim30$-40\% of galaxies brighter than $M_{\rm UV}\lesssim-19$ are in strong upturn phases at $z=14$ in the \textit{increased-burstiness} case, compared to only $\sim20\%$ in the \textit{increased-SFE} scenario. 
The differences between the two models are less pronounced when considering the broader upturn definition ($\rm SFR_3/SFR_{50}>5$). Meanwhile the downturn fractions (bottom panel) show a much stronger separation between the two models. The \textit{increased-SFE} model shows no evolution in the downturn fraction relative to the $z=6$ and 10 fiducial model, as the higher accretion rates at high redshift and the increased stochasticity of lower-mass halos have competing effects that approximately compensate, as discussed in Section~\ref{sec:z6_10_evol}.
In contrast, the \textit{increased-burstiness} model predicts downturn fractions higher by a factor of $\sim2$ relative to $z=6$ and 10 at all magnitudes, reaching $\sim20-30\%$ for galaxies with $M_{\rm UV}\gtrsim-18$. This highlights the importance of detecting galaxies in suppressed phases to constrain burstiness, as their fraction as a function of $M_{\rm UV}$ directly traces the mass dependence of SFR fluctuations $\sigma_{\log\rm SFR}(M_h)$.

At $z=6$ there is already growing evidence for such downturn systems \citep[e.g.][]{Trussler24, Endsley25} and even quiescent phases \citep[e.g. ``lulling'' or ``mini-quenched'' galaxies;][]{Looser23b, Strait23, Looser23}. These sources become increasingly challenging to detect toward higher redshift, as they are expected to be more common at faint UV magnitudes \citep[see also][]{Gelli25a}. However, our models show that, in the case of strongly bursty scenarios, a significant fraction of downturn galaxies is expected even among bright systems at $z=14$, accounting for $\sim20\%$ of galaxies at $M_{\rm UV}\sim-21$. Interestingly, bright galaxies with evidence for recent suppression of star formation have now been observed at $z>9$ \citep{Carniani2025_gsz14, Chen2026, Harikane26}, as we will discuss in Sec.~\ref{sec:discussion}.

Our prediction for the $z=14$ evolution of downturn fractions provides a promising observational test to discriminate between models: an increase of downturn galaxies with redshift, at any fixed $M_{\rm UV}$, would indicate enhanced burstiness, while no evolution would disfavor an increase in the burstiness amplitude.
In the following subsections we turn to more direct observables that are sensitive to different timescales for star formation, to establish empirical tracers that may discriminate between these models.

\subsection{Nebular line equivalent widths}\label{subsec:ew}

Nebular emission line equivalent widths provide a sensitive probe of short-timescale SFH variations: hydrogen recombination lines trace star formation over $\lesssim5$~Myr, while the underlying continuum reflects the accumulated stellar population. Their ratio therefore approximately traces the galaxy's sSFR.
$\rm H\beta$ can be observed with NIRSpec up to $z\sim10$, while $\rm H\gamma$ extends to $z\sim12$. At higher redshifts, $\rm H\alpha$ becomes accessible with MIRI \citep[e.g.][]{Hsiao2024, Helton2025, Zavala2025, Helton2026}. All three lines equivalent widths show similar trends, as they trace the same star-formation timescale. Here, we focus on $\rm H\beta$~EW to facilitate comparison with the available data from \cite{Tang2025} at $z\sim6$-10.

\begin{figure*}[tp]
    \centering
    \includegraphics[width=\linewidth]{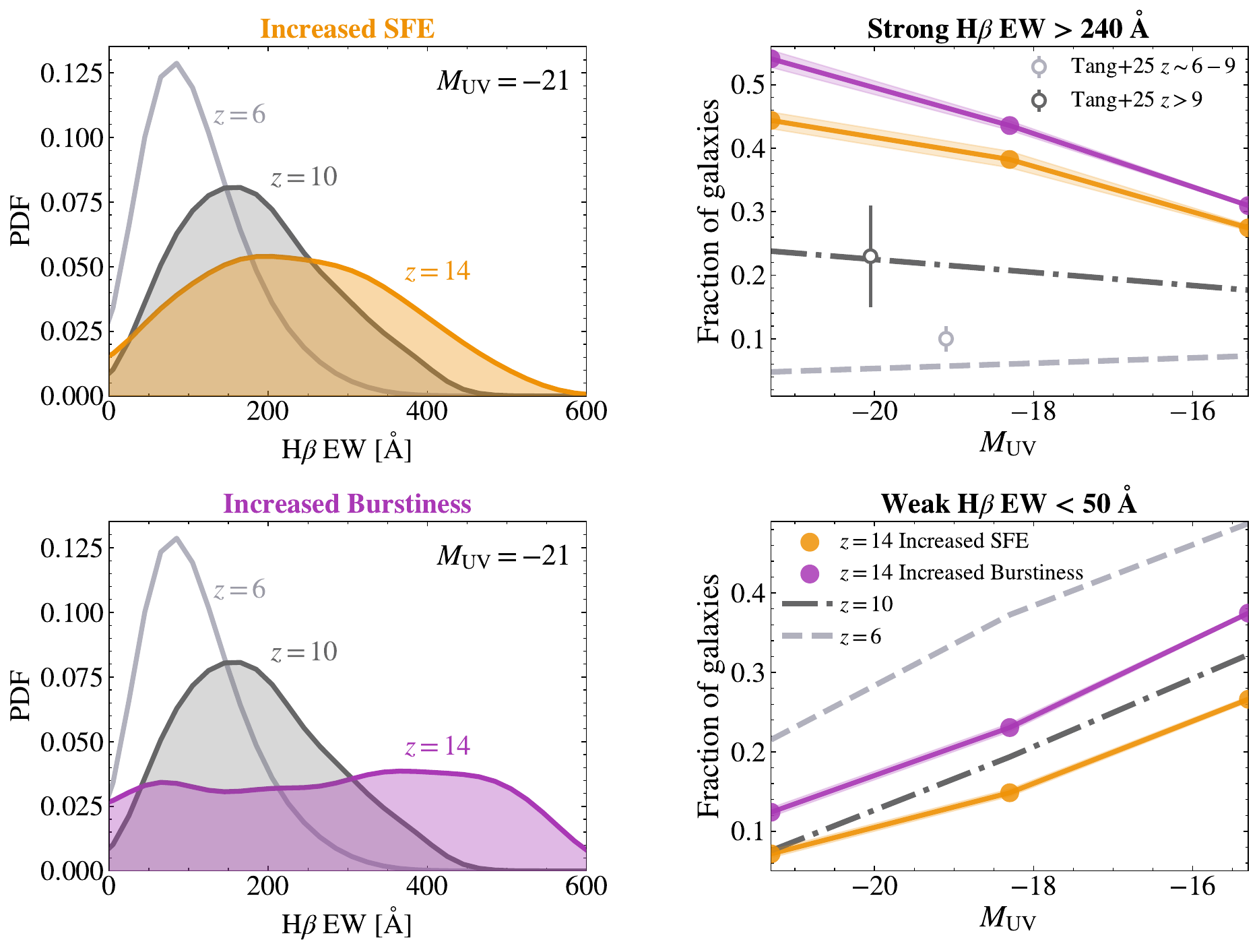}
    \caption{\textit{\textbf{Left:}} Distributions of $\rm H \beta$ EW at fixed $M_{\rm UV}=-21$, for redshifts $z=6$, $z=10$ (fiducial model, in gray) and $z=14$ \textit{increased-SFE} (top) and \textit{increased-burstiness} (bottom) models. The median EW shifts to higher values towards higher redshifts, and the distributions get broader.
    \textit{\textbf{Right:}} Fraction of strong ($\rm H\beta \, EW >240~\angstrom$) and weak ($< 50~\angstrom$) emission line galaxies versus UV magnitude. The strong-line fraction (top) rises from $z=6$ to $z=10$, matching the increase observed by \cite{Tang2025}, and continues rising to $z=14$ in both models, more steeply for \textit{increased-burstiness} scenario. The weak-line fraction (bottom) decreases from $z=6$ to $z=10$ in both models, but diverges at $z=14$: it keeps declining for \textit{increased-SFE}, while it reverses and rises by $\sim1.5\times$ for \textit{increased-burstiness}.
    }
    \label{fig:EW_hbeta}
\end{figure*}

The two panels on the left in Fig.~\ref{fig:EW_hbeta} show the probability distribution functions of $\rm H\beta$ EW at fixed magnitude of $M_{\rm UV}\sim -21$.
We begin by examining the gray distributions (identical in both panels), which illustrate the evolution of the fiducial model from $z=6$ to $z=10$. 
The median EW shifts to higher values toward earlier times, increasing from $\sim 100~\angstrom$ at $z=6$ to $\sim 150~\angstrom$ at $z=10$. 
This increase reflects the higher accretion rates at earlier epochs, which lead to higher typical SFR. At the same time, galaxies at fixed UV luminosity have lower stellar masses, resulting in higher sSFR and therefore larger EW.
The same increase in halo accretion rates toward higher redshift also drives the evolution of the $\rm SFR_3/SFR_{50}$ ratios shown in Fig.~\ref{fig:sfr_ratios}, leading to larger recent-to-past SFR ratios.

Simultaneously, the widths of the distributions are set by the burstiness of star formation.
The width of the EW distribution increases at higher redshift. This EW widening with increasing redshift arises naturally from the mass dependence of burstiness: at fixed $M_{\rm UV}$, galaxies reside in lower-mass halos toward higher-z, and these halos exhibit stronger SFR fluctuations producing broader EW distributions.
As a result, even in the fiducial model at $z=6$ and 10, at fixed $M_{\rm UV}$ we predict both an enhanced high-EW tail (systems observed during recent starbursts) and an enhanced low-EW tail (systems caught in downturn phases) toward higher redshift, even among luminous $M_{\rm UV}\rm -21$ galaxies.

The top-left and bottom-left panels also show the predictions of $\rm H\beta~EW$ distributions for the two $z=14$ star formation models, \textit{increased-SFE} and \textit{increased-burstiness} respectively. In both scenarios, the trends already visible between $z=6$ and $z=10$ continue toward earlier epochs, with the distributions shifting to higher median EWs (due to an increase in accretion rates) and becoming broader (due to mass-dependent burstiness).

However, the models differ in the extent of this broadening. In particular, the \textit{increased-burstiness} scenario produces an extremely wide EW distribution, with both very high-EW ($>500~\angstrom$) and very low-EW ($<20~\angstrom$) galaxies. This is a direct consequence of the large amplitude and time-variability of the SFR. During starburst phases, the SFR can reach typical values higher than in the fiducial and \textit{increased-SFE} models, leading to very high EW, up to $600~\angstrom$ for $M_{\rm UV}\sim -21$ galaxies. Conversely, during downturn phases the SFR can drop to very low values. Given that we adopt variability on timescale of $\sim50$~Myr, typical of SN feedback, the downturns can last longer than the 10~Myr timescale over which the EW of the preceding starburst declines (see Fig.~\ref{fig:single_burst}). As a result galaxies can spend extended periods in very low-SFR state after the starburst signature has faded, leading to extremely low EW. 

To quantify the broadening of the EW distributions, in the right panels of Fig.~\ref{fig:EW_hbeta} we show the fraction of strong and weak-line galaxies, defined as $\rm H\beta\, EW >240~\angstrom$ and $< 50~\angstrom~$ respectively, as a function of $M_{\rm UV}$ and their redshift evolution.
The top-right panel shows the fraction of strong-line galaxies. At fixed UV magnitude, the fraction of strong line emitters increases with redshift because galaxies have higher typical star formation rates at earlier times, shifting the EW distribution toward larger values. This evolution closely mirrors the increase in the fraction of galaxies undergoing recent star formation upturns (Fig.~\ref{fig:sfr_ratios}, left).
The predicted increase of high-EW galaxies with $z$ is consistent with observations by \cite{Tang2025}, who report an increasing fraction of strong emitters from $z=6$ to $z=10$. In particular, for galaxies with $-20<M_{\rm UV}< -19$, they find a fraction of $\sim10\%$ of sources with EW$>240~\angstrom$ at $z\sim6$, rising to $25\%$ at $z\sim10$, in broad agreement with our predicted evolution. The measurement based on the $6<z<9$ sample lies slightly above our $z=6$ prediction, as expected since the observed sample also includes galaxies at higher redshifts, where our model predicts somewhat higher fractions. 

Both the \textit{increased-SFE} and \textit{increased-burstiness} models predict a further increase in the fraction of strong-line galaxies from $z=10$ to $z=14$. However, the increase is substantially larger in the \textit{increased-burstiness} model, reaching $\sim55\%$ for galaxies with $M_{\rm UV}\sim-21$, compared to $\sim45\%$ in the \textit{increased-SFE} model.

The bottom-right panel of Fig.~\ref{fig:EW_hbeta} shows the fractions of galaxies with low $\rm H\beta$~EW.
At all redshifts, we predict a higher fraction of weak-line systems among fainter galaxies, as the faint end is more strongly populated by lower-mass galaxies undergoing downturns.
From $z=6$ to $z=10$, the fraction of weak-line galaxies decreases by a factor of $\sim2$ at all magnitudes, because the median $\rm H\beta~EW$ is higher at earlier epochs. 
However, the behavior at $z=14$ depends strongly on the assumed star formation model. In particular, the \textit{increased-burstiness} scenario yields higher fraction than at $z\sim10$, whereas the \textit{increased-SFE} model predicts a further decline. In other words, if burstiness increases toward earlier times, the fraction of low-EW galaxies is expected to increase from $z=10$ to 14.

Measuring the fractions of strong and weak line emitters and their redshift evolution would provide a powerful way to break degeneracies between models. In particular, an increasing population of weak line emitters at fixed $M_{\rm UV}$ with redshift is a distinctive signature of increasingly bursty star formation histories.

\subsection{Balmer break}\label{subsec:balmer_break}

Having examined how bursty star formation histories shape the nebular emission line EWs, which trace star formation over the past $\lesssim 5$~Myr, we now turn to the Balmer break. This provides a complementary test of our models by tracing the emergence of spectra dominated by older ($\gtrsim20$-30 Myr) stellar populations.

\begin{figure}
    \centering
    \includegraphics[width=\linewidth]{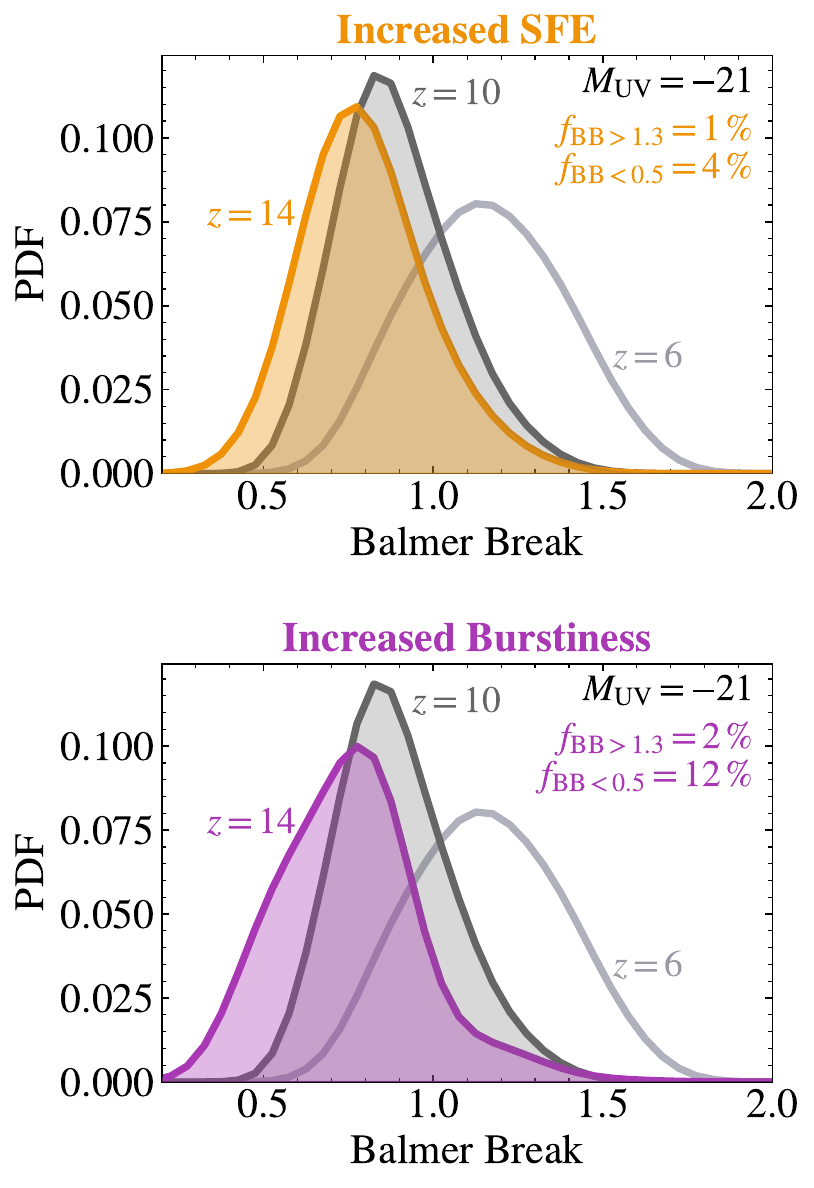}
    \caption{Distributions of Balmer breaks ($F_\nu(4200\angstrom)/F_\nu(3500\angstrom)$) at fixed $M_{\rm UV}\sim-21$, for redshifts $z=6$, $z=10$ (fiducial model, in gray) and $z=14$ \textit{increased-SFE} (top) and \textit{increased-burstiness} (bottom) models. The median Balmer break shifts to lower values towards higher redshifts, and at $z=14$ the distribution is broader in the \textit{increased-burstiness} compared to \textit{increased-SFE}, as highlighted by the reported fractions of weak and strong BB galaxies.}
    \label{fig:balmer_break}
\end{figure}

Fig.~\ref{fig:balmer_break} illustrates the distributions of Balmer break strengths at fixed magnitude $M_{\rm UV}\sim -21$.
From redshift $z=6$ to $z=10$ the typical median Balmer break values become lower. This trend is expected, as galaxies at higher redshift are on average found at earlier stages of their assembly and therefore host younger stellar populations with fewer old stars.

Focusing on the different $z=14$ star formation models in the two panels, the \textit{increased-SFE} model shows no significant evolution in the shape of the Balmer break strength distribution with respect to $z=10$, with only a shift to lower values due to the average younger stellar ages.
The scenario with \textit{increased-burstiness} produces a broader distribution with respect to $z=10$, spanning both very low and very high Balmer break values, from $0.3$ to $1.6$.
In particular, galaxies with strong Balmer jumps ($\rm Balmer~break<0.5$) represent 12\% of the \textit{increased-burstiness} population, compared to only 4\% in the \textit{increased-SFE} case.
Very strong ($>1.3$) Balmer break galaxies correspond instead to galaxies dominated by older stellar populations ($\gtrsim30$~Myr, see Fig.~\ref{fig:single_burst}) from previous starburst phases. These are very rare in both scenarios at $z=14$, but twice more common in the \textit{increased-burstiness} model (2\%) compared to the \textit{increased-SFE} (1\%).
Detections of Balmer breaks at such high redshift would therefore provide a signature of strong burstiness in star formation histories at these epochs. Such observations are feasible thanks to the JWST/MIRI instrument, which can access the continuum redward of the Balmer break for $z>10$ galaxies \citep[e.g.][]{Helton2025}. Recent examples of Balmer break detections are now reported \citep[e.g.][]{Harikane26}, providing direct evidence of evolved stellar populations that may have formed during earlier starburst episodes, highlighting the power of this diagnostic to test early star formation histories.

\subsection{UV slope}\label{subsubsec:uv_slopes}

\begin{figure}
    \centering
    \includegraphics[width=\linewidth]{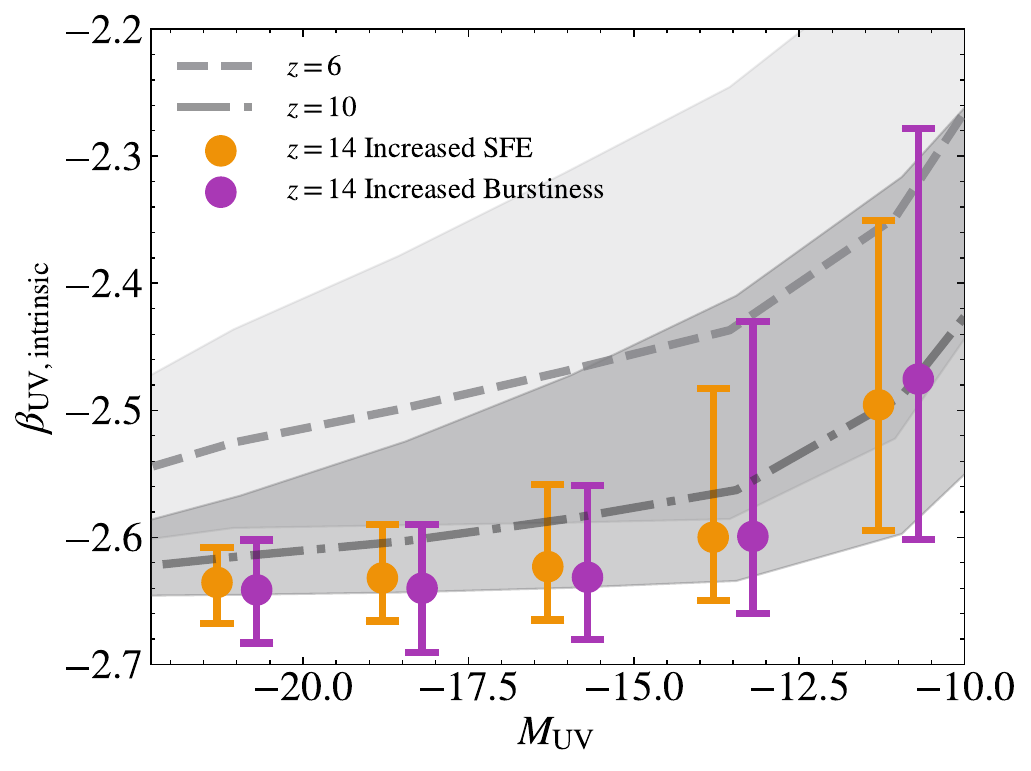}
    \caption{Median intrinsic $\beta_{\rm UV}$ vs $M_{\rm UV}$ at $z=6$, 10, and 14 (both $z=14$ models). Galaxies become progressively bluer at fixed $M_{\rm UV}$ toward higher redshift, driven by increasing accretion rates and younger stellar populations. At $z=14$, \textit{increased-burstiness} predicts more very blue, bright galaxies ($\beta_{\rm UV}\lesssim-2.7$ at $M_{\rm UV}<-18$) and more very red, faint galaxies ($\beta_{\rm UV}\gtrsim-2.35$ at $-15<M_{\rm UV}$) than \textit{increased-SFE}.
    }
    \label{fig:UV_slope}
\end{figure}

We now focus on the UV continuum slope $\beta_{\rm UV}$, which responds to variations in the SFH on up to $\sim100$~Myr timescales. Following a burst, $\beta_{\rm UV}$ becomes redder as the burst fades, with variations of $\Delta\beta_{\rm UV}\sim0.5$ over $\sim100$~Myr (see Fig.~\ref{fig:single_burst}). As in Section~\ref{Sec:fiducial} we show the \textit{intrinsic} UV slopes, before dust attenuation, as $z\gtrsim9$ galaxies appear to be relatively unattenuated \citep[e.g.][]{Cullen2024, Morales2024, Topping2024}.
We focus in particular on the redshift evolution of the $\beta_{\rm UV}$--$M_{\rm UV}$ relation.
Fig.~\ref{fig:UV_slope} shows the trend for $z=6$, 10 and 14, using the same color coding of Fig.~\ref{fig:sfr_ratios} for the two $z=14$ star formation models.
The redshift evolution reveals that galaxies at all magnitudes become progressively bluer towards earlier epochs.
The general trend of ${\beta_{\rm UV}}$ decreasing with $z$ is commonly observed \citep[e.g.][]{Saxena2024, Topping2024} and usually attributed to the presence of dust which, building up at later times and in more evolved galaxies, plays a great role in reddening the spectra at lower $z$. However, here we show that, even when not including dust, spectra are expected to be bluer at high-$z$. Galaxies with $M_{\rm UV}=-20$ go from median intrinsic UV slopes of $\sim-2.5$ at $z\sim6$ to $\sim-2.65$ at $z\sim14$.
This evolution is primarily driven by the increasing halo accretion rates at early times, which lead to higher average SFR and systematically younger stellar populations. 

At $z=14$, the two star-formation models both predict very blue UV slopes on average, $\beta_{\rm UV}\sim-2.65$ for $M_{\rm UV}\sim-20$, reflecting the extremely young ages of the stellar populations. However, the \textit{increased-burstiness} model tends to have broader distributions of possible $\beta_{\rm UV}$ values at all magnitudes, including a substantial population of faint galaxies (30\% in the range $-15<M_{\rm UV}<-10$) with intrinsic UV slope ${\beta_{\rm UV}}\gtrsim-2.4$, compared to 20\% in the increased-SFE model. These systems correspond to galaxies undergoing downturn phases in their star formation histories, with the last strong burst having occurred more than $\sim40$~Myr earlier. In practice, detecting such faint sources at $z>10$ and robustly measuring their UV slopes is extremely challenging even for JWST, although recent observations are pushing the current limits, with reported examples of red sources at $z>10$ \citep[e.g.][]{Donnan2025_capers}.
Differences between the two $z=14$ models also arise at the bright end, where the \textit{increased-burstiness} model predicts a larger fraction of very blue galaxies with $\beta_{\rm UV}<-2.7$ at $M_{\rm UV}<-20$: 9\% compared to only 2\% in the \textit{increased-SFE} model. We recall that these predictions do not include dust attenuation, which would likely redden the galaxies, particularly the more evolved and typically brighter ones. Nevertheless, an excess of very blue galaxies at the bright end together with very red galaxies at the faint end could hint at enhanced burstiness at high redshift.

\subsection{$z>15$ luminosity functions}\label{subsec:z20uvlf}

At the time of writing, the highest-redshift spectroscopically confirmed galaxy lies at $z=14.44$ \citep{Naidu2026_MoM}, with $M_{\rm UV} = -20.2$. While several galaxy candidates have been reported at even higher redshifts ($15 < z < 20$; e.g. \citealt{Castellano2025, Kokorev2025_glimpsehighz, Perez-Gonzalez2025}), none have yet been spectroscopically confirmed.

The confirmation of galaxies at these extreme epochs, or non-detections providing more stringent upper limits on the UVLF, may provide additional constraints on star formation models. In particular, models that are calibrated to reproduce the UVLF at $z\sim14$ may begin to diverge significantly at higher redshifts, offering a way to break degeneracies.

\begin{figure}[bp]
    \centering
    \includegraphics[width=\linewidth]{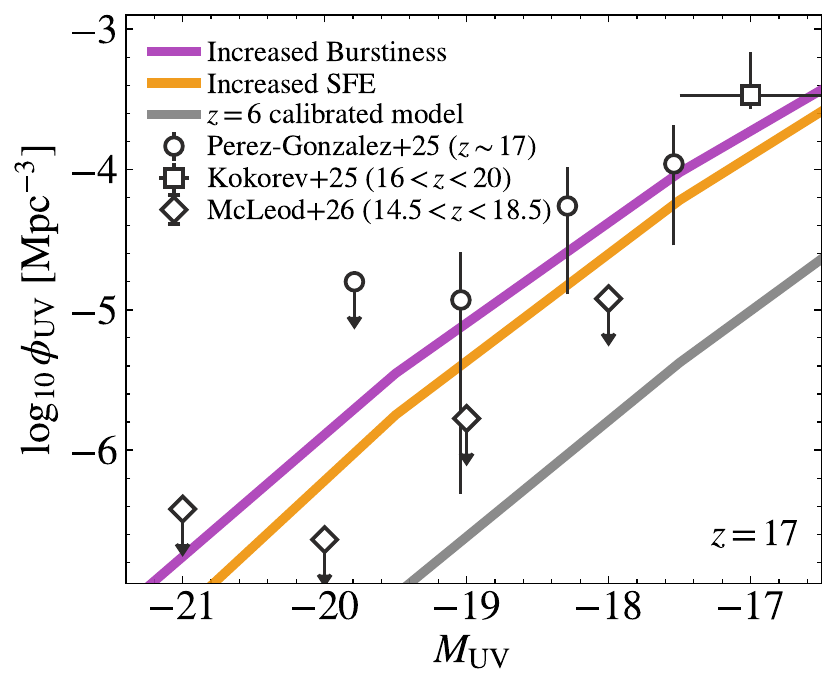} 
    \caption{Predictions of $z=17$ UVLF for the \textit{increased-burstiness}, \textit{increased-SFE} and fiducial models, compared with data from \cite{Kokorev2025_glimpsehighz, Perez-Gonzalez2025, McLeod2026}.
    Both $z=14$ models predict number densities $>10\times$ higher than the fiducial $z=6$ calibrated model, with \textit{increased-burstiness} systematically higher than \textit{increased-SFE} due to the growing contribution of more bursty low-mass halos.
    }
    \label{fig:z20uvlf}
\end{figure}

To explore this, in Fig.~\ref{fig:z20uvlf} we show the $z\sim17$ UVLFs predicted by the fiducial model calibrated at $z=6$ (gray curve) and the two $z=14$ models, i.e. \textit{increased-burstiness} and \textit{increased-SFE}. Both $z=14$ models predict higher number densities at $z=17$ than the fiducial model by a factor $>10$.
The \textit{increased-burstiness} model predicts higher number densities compared to the \textit{increased-SFE} scenario. This difference arises from the growing contribution of low-mass halos, which are both more abundant and more strongly affected by burstiness due to the mass-dependent stochasticity.
As discussed in Section~\ref{sec:highz} and in \cite{Gelli24b}, this effect becomes increasingly important at higher redshift: the combination of the steep halo mass function and $\sigma_{\log \mathrm{SFR}}(M_h)$ decreasing with $M_h$ boosts the relative contribution of the numerous low-mass systems to the UVLF. 

In the figure, we also show current photometric constraints on the $z\sim17$ UVLF. Results from \cite{Perez-Gonzalez2025} and \cite{Kokorev2025_glimpsehighz} show an excess of galaxy candidates relative to pre-JWST predictions at $z=17$, suggesting that the overabundance observed at $z=14$ may persist at higher redshifts.
While \cite{Kokorev2025_glimpsehighz} data seem to tentatively favor the \textit{increased-burstiness} model, the current uncertainties cannot robustly rule out either scenario. 
\cite{McLeod2026} instead finds number densities that lie below constraints from other data, indicating a more rapid dearth of galaxies above $z>15$, potentially disfavoring both the \textit{increased-burstiness} and \textit{increased-SFE} scenarios as currently calibrated. However, the inferred upper limits on the galaxy number density remain approximately one order of magnitude above the $z=17$ fiducial prediction, so we cannot rule out that a modification to the star formation histories may still be required at $z=17$.

At present, the observational picture at $z\gtrsim15$ remains highly uncertain, with different analyses reaching substantially different inferred abundances. This regime is based on photometric candidates and small-number statistics, with a limited number of independent deep fields, making the inferred abundances particularly sensitive to contamination and selection effects and cosmic variance.
Larger samples and confirming the nature of the $z>14.5$ candidates will therefore be crucial to more reliably establish the $z>14.5$ population and constrain star formation models.

\section{Discussion}\label{sec:discussion}

Our results show that halo mass-dependent burstiness in star formation rate is essential to reproduce the observed properties of the $z=6$ galaxy population and their evolution to $z=10$, in particular the fractions of galaxies undergoing star-formation upturns and downturns.
At $z=14$ we showed that scenarios with \textit{increased-burstiness} and \textit{increased-SFE}, both built to reproduce the $z=14$ UVLF observed by JWST, predict distinct distributions of spectral observables, offering a path to interpret the spectral diversity observed in JWST data and break the UVLF degeneracy.
We now explore the most promising observational avenues to constrain stochasticity using galaxies in star-formation downturn phases (Sec.~\ref{disc:downturns}) and discuss future prospects (Sec.~\ref{disc:future}).

\subsection{Constraining stochasticity with star formation downturns}\label{disc:downturns}

Our results show that the fraction of galaxies caught in star formation downturns provides one of the most promising probe of the stochasticity of star formation (Sec.~\ref{subsec:sfr_ratio},\ref{subsec:ew}). In particular, the dependence of the downturn fraction on $M_{\rm UV}$ reflects the mass dependence of the stochasticity, while its overall amplitude traces the strength of the fluctuations. Measuring the downturn fraction over a range of $M_{\rm UV}$ and redshifts can therefore constrain how stochasticity and its mass dependence evolve with cosmic time.
The fraction of galaxies in recent upturns is similarly sensitive to stochasticity, but its evolution with redshift is degenerate with the increasing average gas accretion rate. Downturns are less subject to this degeneracy and therefore provide a cleaner probe of stochasticity.

\begin{figure}
    \centering  
    \includegraphics[width=0.95\linewidth]{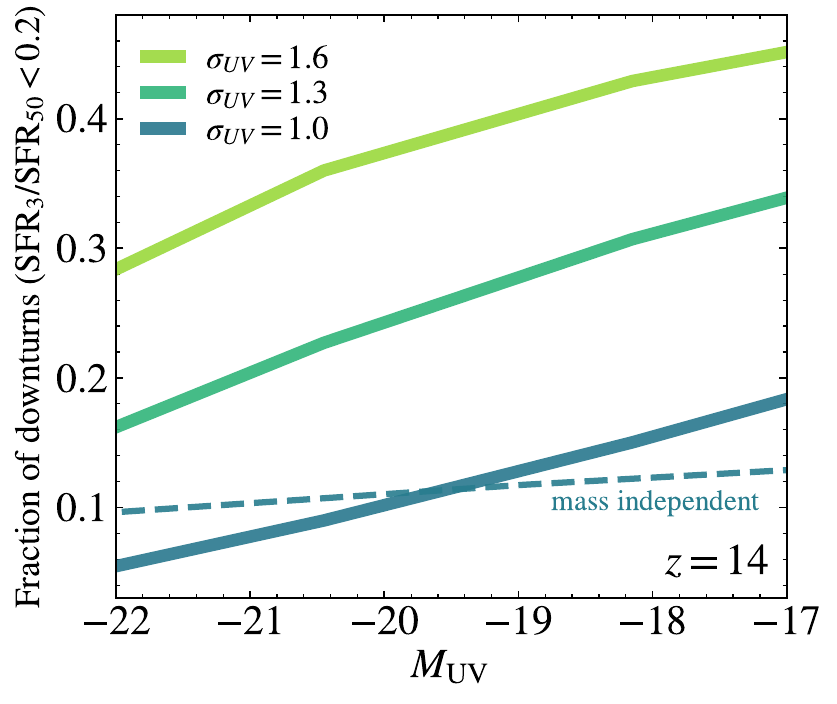}
    \caption{Fraction of galaxies in downturns, with $\rm SFR_3/SFR_{50}<0.2$, as a function of $M_{\rm UV}$ for models with increasing stochasticity of star formation ($\sigma_{\rm UV}(M_h=10^{10}\msun)=1$, 1.3, 1.6).The dashed line shows a model with mass-independent scatter and $\sigma_{\rm UV}(M_h=10^{10}{\rm M_\odot}) = 1$~mag. The downturn fraction at fixed $M_{\rm UV}$ increases with the burstiness level, and rises toward faint $M_{\rm UV}$ only when the scatter is mass-dependent.}
    \label{fig:z14downturn_sigmaUV}
\end{figure}

To illustrate how downturn phases, here defined as $\rm SFR_3/SFR_{50}<0.2$, directly correlate with burstiness, we realize a set of models at $z=14$ with increasingly high amplitude of bursts (scaling star formation efficiency to match the observed UVLF). In Fig.~\ref{fig:z14downturn_sigmaUV} we show the fraction of galaxies in downturns as a function of the UV magnitude for models spanning a range of stochasticity, quantified by the UV scatter at $M_h = 10 ^{10}\msun$.
Because the scatter is mass-dependent, downturns are more common at fainter $M_{\rm UV}$ in every model. For comparison, a model with mass independent scatter and $\sigma_{\rm UV}(M_h=10^{10}\,\msun)=1$ is also shown, highlighting that in this case the fraction of downturns does not vary with $M_{\rm UV}$.
Crucially, at fixed $M_{\rm UV}=-21$ the downturn fraction increases by a factor of $\sim4$ as $\sigma_{\rm UV}(M_h=10^{10}\,\msun)$ increases from $1$ to $1.6\,\rm mag$. In the most extreme model ($\sigma_{\rm UV}=1.6\,\rm mag$), more than 35\% of $M_{\rm UV}=-20$ galaxies are caught in a downturn at any given time. The downturn fraction at fixed $M_{\rm UV}$ is thus a direct, quantitative estimator of burstiness.

Measuring the downturn fraction observationally requires robustly recovering the SFH of high redshift galaxies.
As shown in Section~\ref{subsec:ew} and Fig.~\ref{fig:EW_hbeta}, weak recombination line EWs are a key tracer of downturns: the fraction of weak-EW systems increases by $\sim1.5\times$ from $z=10$ to $z=14$ in the \textit{increased-burstiness} model, while it declines over the same interval in the \textit{increased-SFE} model. Although the predicted change is relatively small, the opposite evolution provides a clear qualitative distinction between the two scenarios.
However, low EWs are not a unique signature of suppressed star formation as they could also arise when high fractions of ionizing photons escape the galaxy \citep[e.g.][]{Endsley23b}. 
Ongoing and upcoming JWST programs provide promising opportunities to break this degeneracy. Deep NIRSpec grating spectroscopy enables more accurate reconstruction of star formation histories through resolved low-ionization absorption lines that can provide priors on the ionizing photon escape fraction $f_{\rm esc}$, breaking degeneracies present in prism spectra \citep[e.g.][]{Chen2026, Vasan2026}.

Recovering $\rm SFR_3/SFR_{50}$ directly would be an even stronger diagnostic for constraining stochasticity and distinguishing $z>10$ models, but requires combining emission-line EWs with information on the UV continuum and Balmer break strengths. In this respect, MIRI observations can probe Balmer breaks and H$\alpha$ emission at $z>7$, providing complementary diagnostics for both recent and older star formation activity \citep[e.g.][]{Hsiao2024miri, Helton2025, Helton2026, Harikane26}.

Intriguingly, recent observations are beginning to provide examples of bright high-redshift systems consistent with galaxies in post-burst downturn phases.
\citet{Chen2026} report a $z=9.3$ $M_{\rm UV} = -21.66$ source with a  Balmer break and low H$\beta$ EW of $\sim25\angstrom$, and weak UV emission and stellar wind features, consistent with a galaxy in a downturn phase in the past $10-20$~Myr. 
Similarly, a recent study by \cite{Harikane26} reported the detection of a Balmer break ($>1.1$) in a $M_{\rm UV}\sim -20$ galaxy at $z=11$. The detected Balmer break, together with the lack of strong H$\alpha$ and [OIII] emission lines, implies a star formation history in which this galaxy experienced an intense burst of star formation approximately $\sim 10$ Myr prior to the time of observation, followed by a phase of suppressed activity.
Current samples with deep grating resolution spectroscopy or MIRI coverage that enable robust estimates of past star-formation are still small, but these detections show that downturn phases are already identifiable at $z>9$.

Expanding these samples to determine the fraction of $z>10$ galaxies in downturn phases, as a function of $M_{\rm UV}$ and redshift, will be important for constraining the role of stochastic star formation at high redshifts. In particular, if increased burstiness is responsible for the high abundance of luminous galaxies at $z\sim14$, our model predicts that downturns become more common with increasing redshift at fixed $M_{\rm UV}$.

\subsection{Future prospects}\label{disc:future}

Our framework provides a flexible and predictive approach to exploring a wide range of bursty star formation models and the signatures they leave on SEDs.
Here, we discuss several physical ingredients that can impact bursty star formation and consider which assumptions could be relaxed to further explore their role and identify additional observables.

Despite the success of our fiducial mass-dependent model at $z\sim6-10$, departures from observations point toward important additional physics shaping galaxies' spectra.
In particular, in Section~\ref{Sec:fiducial} we showed that the model underproduces the downturn fraction in the faintest magnitude bin reported by \citet{Endsley25} by a factor $\sim1.5\times$.
The higher observed downturn fraction  could point to suppressed phases lasting longer in low-mass galaxies than our model currently allows. 
In our model, we characterize SFH fluctuations with a single timescale of $\tau\sim50$ Myr, consistent with values found in high-redshift galaxy simulations. 
However, multiple physical processes operating over different timescales could in principle give rise to more complex SFHs characterized by multiple timescales \citep[e.g.][]{Tacchella20}. 
In addition, the duty cycle of starbursts and downturns may not be symmetric, particularly in low-mass galaxies where strong feedback episodes can easily evacuate gas and delay the recovery of star formation \citep[e.g.][]{Furlanetto22, Dome24, Gelli24a, Gelli25a}. 
Incorporating these refinements could resolve the faint-end discrepancy with \citet{Endsley25}. However, because this effect is again driven by a halo-mass dependence, it is not expected to alter the redshift evolution of the downturn fraction, which retains a similar slope across epochs (Sec.~\ref{subsec:sfr_ratio}). This highlights the importance of constraining the luminosity dependence of downturn fractions, which can also provide key information on the characteristic timescales and duty cycles governing stochastic star formation.

The faint-end downturn discrepancy could also reflect a higher ionizing photon escape fraction in the faintest galaxies, which would suppress the inferred EW independently of the underlying SFH, however we expect this may be a sub-dominant effect to SFH.
We do not explicitly model ionizing photon escape fractions or their temporal variability. Variations in $f_{\rm esc}$ could affect observables such as $\beta_{\rm UV}$ and nebular emission line equivalent widths, especially immediately following starburst episodes. In particular, high escape fractions can temporarily produce very blue UV slopes over short timescales ($<5$ Myr; \citealt{Topping2024}), introducing additional degeneracies with SFH-driven signatures. 
However, this effect is expected to impact only a small fraction of galaxies: as shown by \citet{Topping2024}, very blue galaxies ($\beta_{\rm UV}<-2.8$), which require very high escape fractions to explain, account only for $\lesssim4\%$ of their JADES sample of galaxies at $z\gtrsim6$ with $M_{\rm UV}>-20$.
Deep JWST spectroscopy providing detections of multiple low-ionization absorption features may independently place constraints on $f_{\rm esc}$ in high-redshift galaxies \citep[e.g.][]{Chen2026, Vasan2026}, helping to disentangle these effects.

Our results show that increased burstiness could provide an explanation for the diverse spectral properties of galaxies at $z>10$, raising the question of which physical processes can drive such variation in SFR and suggesting strong feedback processes at play. One interesting possibility could be variation of the stellar IMF, as a larger population of massive stars could, in addition to increasing the UV luminosity produced per unit stellar mass \citep[e.g.][]{Rasmussen23, Trinca23, Hutter2025}, also enhance stellar feedback, and give rise to more bursty star formation histories \citep[e.g.][]{Cueto2026}. 
Recent ultra-deep rest-UV spectroscopy of $z\gtrsim9$ galaxies, including GN-z11, has revealed stellar-wind features consistent with a contribution from very massive stars with masses $>100~\msun$ \citep{Chen2026_gnz11, Marques-Chaves2026}.
From a theoretical perspective, the IMF may depend on the physical conditions of star formation, with studies suggesting that low-metallicity environments could favor a more top-heavy IMF \citep[e.g.][]{Chon2024}.
Our model currently assumes a Kroupa IMF with an upper mass limit of $100~\msun$, and therefore does not directly capture these effects. 
Future work exploring IMF variations and the contribution of very massive stars and their feedback will be important to understand the interplay with bursty star formation and the resulting spectral signatures at $z>10$.

Beyond the stellar emission and spectral features explored here, bursty star formation may also affect other observable properties.
In particular, bursty star formation may impact gas-phase metallicities,
as episodes of intense star formation may drive both rapid metal production and feedback-driven outflows. 
This can introduce scatter in the mass-metallicity relation \citep[e.g.][]{Curti23, Pallottini24}. Although metallicity variations are expected to have a smaller impact on $M_{\rm UV}$ than variations in SFR, they could strongly affect metal emission-line strengths and ratios, providing additional observables with which to test star formation stochasticity. Our model assumes a fixed metallicity of $Z_\star=0.1\rm~Z_\odot$, but in future work, exploring the interplay between metallicity and bursty star formation could provide additional constraints on the physical processes shaping early galaxies.

Overall, our results show that spectral diagnostics provide a powerful avenue to distinguish between different physical scenarios for early galaxy formation. Combining this approach with additional probes, such as metallicity-sensitive observables and galaxy clustering \citep[see e.g.,][]{Munoz23, Gelli24b, Nikoli2026, Sun2026}, will provide complementary constraints on the physical processes shaping the first galaxies.

\section{Conclusions}\label{sec:conclusions}

We developed a framework to model bursty star formation histories across cosmic time, with the goal of assessing the impact of stochasticity on multiple galaxy observables. We model stochastic fluctuations of the star formation histories of galaxies, characterized by the halo mass-dependent instantaneous star formation efficiency $\epsilon_\star(M_h)$ and the burst amplitude $\sigma_{\log\rm SFR}(M_h)$. 
We generate galaxy populations by sampling halos from the halo mass function at each redshift and assigning bursty star formation histories, which we then forward-model into spectral energy distributions to predict key observables ($\beta_{\rm UV}$, hydrogen emission-line EW, and Balmer break strengths).
We calibrate the model at $z=6$ to reproduce both the UV luminosity function and the observed ratios of $\rm SFR_3/SFR_{50}$.
Our main findings are:
\begin{itemize}
    \item A significant level of burstiness is required at $z=6$ to reproduce the distributions of $\rm SFR_3/SFR_{50}$ inferred from UV-to-optical SED fitting by \cite{Endsley25}, corresponding to a UV scatter of $\sigma_{\rm UV}(10^{10}\,\rm M_\odot) = 0.9\,mag$. \\
        
    \item The burstiness must be \textit{mass-dependent}, with low-mass halos more bursty than high-mass ones, to reproduce the increased fractions of galaxies in downturn phases ($\rm SFR_3/SFR_{50}<0.2$) toward fainter $M_{\rm UV}$ at $z=6$ observed by \citet{Endsley25}. The same model also reproduces the broader distributions of $\rm H \alpha$~EW toward fainter $M_{\rm UV}$ observed by \citet{Endsley23b} and the evolution in upturn and downturn fractions from $z=6$ to $z=10$ inferred from NIRSpec spectra by \cite{Tang2025}, which a mass-independent model fails to reproduce. 
    This mass dependence is consistent with that found by \cite{Munoz2026}, and reflects the greater susceptibility of lower-mass halos to star formation suppression. \\

    \item When extrapolated from $z=6$ to higher redshifts, at fixed $M_{\rm UV}$, the model predicts that the median observables evolve toward: higher $\rm H\beta~EW$, bluer UV slopes, and weaker Balmer breaks, mainly driven by increasing halo accretion rates. Simultaneously, the distributions of $\rm SFR_3/SFR_{50}$ and $\rm H\beta~EW$ broaden toward high redshift, as galaxies at fixed $M_{\rm UV}$ reside in lower-mass, more bursty halos. \\
    
    \item Our mass-dependent model predicts that the fraction of galaxies in upturns ($\rm SFR_3/SFR_{50}>5$) at fixed $M_{\rm UV}$ naturally rises with redshift, with a larger increase at brighter magnitudes. This is in agreement with the increase from 10\% to 25\% between $z\sim6$ and 10 estimated by \cite{Tang2025} from JWST spectroscopy. Thus, we predict the bright end of the UVLF becomes increasingly dominated by galaxies undergoing recent star formation upturns at higher redshift.\\

    \item The mass-dependent model predicts that the downturn fraction ($\rm SFR_3/SFR_{50}<0.2$) rises toward fainter $M_{\rm UV}$, directly reflecting the mass dependence of burstiness, but remains nearly constant with redshift unless stochasticity increases.\\

\end{itemize}

When extrapolated beyond the $z=6$ calibration, the mass-dependent model remains in very good agreement with the observed UVLF up to $z=10-11$. However, it starts to fall short at $z\gtrsim12$ at the bright end, and by $z=14$ it underpredicts the observed abundance of UV-luminous galaxies by $\sim1.5$~dex, demonstrating that the level of mass-dependent burstiness inferred from the $z=6$ calibration is insufficient to explain the abundance of UV-luminous galaxies at the earliest epochs.

Motivated by this discrepancy between the observed JWST $z\sim14$ UVLF and the predictions from the $z=6$-calibrated model, we explore two modifications to the star formation histories that reproduce the observed $z=14$ UVLF: \textit{increased burstiness} ($\sigma_{\rm UV}(M_h=10^{10}\msun) = 1.4\rm\,mag$) and \textit{increased star formation efficiency} (a factor $\times5$ in the $\epsilon_{\star,0}$ parameter). While these two scenarios are fully degenerate in their predicted number densities, they produce distinct and testable signatures in the distributions of spectral observables:

\begin{itemize}

    \item The abundance of galaxies undergoing a recent downturn of star formation strongly depends on the level of burstiness, with downturns becoming more common toward fainter $M_{\rm UV}$ as a consequence of the mass dependence of burstiness.
    In particular, in the fiducial and \textit{increased-SFE} model, the fraction of systems with $\rm SFR_3/SFR_{50}<0.2$ remains approximately constant from $z=6$ to $z=14$ at all $M_{\rm UV}$, while it increases by a factor of $\sim2$ in the \textit{increased-burstiness} scenario, as a consequence of the more common suppression phases in the presence of strong stochasticity; \\

    \item Stronger SFR variations during galaxy evolution cause the \textit{increased-burstiness} model to produce significantly broader distributions of hydrogen emission line equivalent widths, UV slopes and Balmer breaks compared to the \textit{increased-SFE} model; \\

    \item The \textit{increased-burstiness} model predicts an increasing population of low-equivalent-width galaxies due to more frequent strong downturn phases: the fraction of galaxies with $\rm H\beta~EW<50~\angstrom$ is expected to increase by a factor of $\sim1.5$ from $z\sim10$ to $z\sim14$ at $-22<M_{\rm UV}<-18$, while it decreases in the \textit{increased-SFE} scenario. \\
    
\end{itemize}

Overall, our results show how the striking spectral diversity observed among UV-luminous galaxies at $z > 10$ is a natural consequence of bursty star formation histories, and that its statistical characterization provides a powerful way to constrain models of early galaxy formation. In particular, we predict that the abundance of galaxies in downturn phases represents a key discriminant between competing scenarios. While individual post-burst systems have already been identified at $z > 9$ \citep{Tang2025, Chen2026, Harikane26}, building statistically significant samples and tracking their abundance as a function of redshift and UV magnitude will be key for distinguishing enhanced burstiness from increased star formation efficiency models at $z\sim14$. Deep NIRSpec spectroscopy, combined with MIRI follow-up to access Balmer breaks and H$\alpha$, makes this an achievable goal in the near future.

\begin{acknowledgments}
We thank Ryan Endsley for providing useful comments on a draft of the manuscript, and Robert Feldmann and Andrei Mesinger for useful discussions.
VG acknowledges support by the Carlsberg Foundation (CF22-1322).
CAM acknowledges support by the European Union ERC grant RISES (101163035), Carlsberg Foundation (CF22-1322), and VILLUM FONDEN (37459). 
JBM was supported by the National Science Foundation under Grants AST-2307354 and AST-2408637, and NASA through grant JWST-GO-03224.
Views and opinions expressed are those of the author(s) only and do not necessarily reflect those of the European Union or the European Research Council. Neither the European Union nor the granting authority can be held responsible for them.
\end{acknowledgments}

\restartappendixnumbering
\appendix

\section{UV magnitude distributions}\label{Appendix:pdf_Muv}

The relation between the halo mass and UV luminosity, commonly described through the conditional distribution $P(M_{\rm UV}|M_h)$, is a key ingredient in galaxy formation models for connecting UV luminosity functions to the underlying dark matter halo population.

In Section~\ref{Sec:fiducial}, we find a scatter of the relation of $\sigma_{\rm UV}(M_h=10^{10}\msun)=0.9\rm~mag$ in our fiducial model, slightly lower than the value inferred by \cite{Gelli24b}. However, in \cite{Gelli24b}, the conditional distribution $P(M_{\rm UV}|M_h)$ is assumed to be Gaussian, as done in many empirical and analytical models, whereas in our current framework it is not imposed a priori and instead emerges self-consistently from the stochastic star formation histories, and can therefore deviate from a Gaussian. Since $P(M_{\rm UV}|M_h)$ is key in determining how galaxies are redistributed across the UV luminosity function, here we examine the shape of the distribution as predicted by our model.

\begin{figure}
    \centering
    \includegraphics[width=0.43\textwidth]{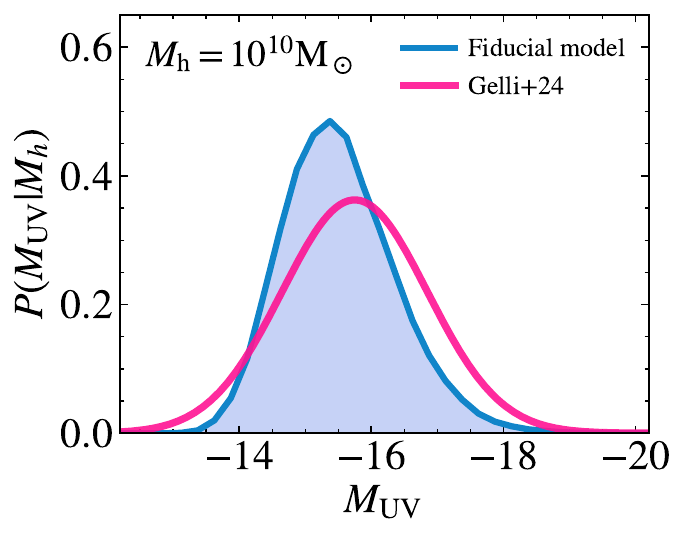}
    \caption{Conditional distribution of UV magnitude $P(M_{\rm UV}|M_h)$ at $M_h = 10^{10}{\rm M_\odot}$ and $z=6$. The blue distribution shows our fiducial model, compared with the gaussian distribution assumed in \cite{Gelli24b} (pink). The stochastic SFH model predicts a clearly non-gaussian distribution, skewed toward higher luminosities relative to the gaussian case.
    }
    \label{fig:pdf_muv}
\end{figure}

Figure~\ref{fig:pdf_muv} shows the resulting distribution for halos with $M_h=10^{10}\msun$, compared to the Gaussian approximation adopted in \cite{Gelli24b}. The model distribution is clearly non-Gaussian, with a pronounced skewness toward higher luminosities. Thus, while $\sigma_{\rm UV}=0.9\rm~mag$ provides a useful measure of its overall width, it does not fully characterize the effect of stochasticity on the UV luminosity function.
This asymmetry arises naturally from the non-linear mapping between bursty star formation histories and UV emission. Following a burst, the UV luminosity declines gradually over tens of Myr (Fig.~\ref{fig:single_burst}), so galaxies spend an extended period in moderately UV-bright states. Conversely, the suppression of UV luminosity during downturn phases is limited because stars formed in previous episodes continue to contribute to the UV continuum. The observed UV emission therefore reflects the superposition of stellar populations with different masses and ages, producing an intrinsically asymmetric distribution.
Such behavior is also found in high-resolution simulations that explicitly resolve bursty star formation \citep[e.g.][]{Ma18, Katz23, Sun23b}, indicating that it is a generic consequence of stochastic SFHs rather than a feature specific to our model.

The skewness of $P(M_{\rm UV}|M_h)$ therefore modifies the impact of scatter on the UV luminosity function relative to the standard Gaussian scatter typically assumed in analytical and empirical models. The relative contribution of up-scattered and down-scattered halos differs from the Gaussian case, affecting both the normalization and the shape of the UV luminosity function.

\section{Redshift \lowercase{$z=10-13$} UVLF}\label{Appendix:z10UVLF}

\begin{figure*}
    \centering
    \includegraphics[width=\linewidth]{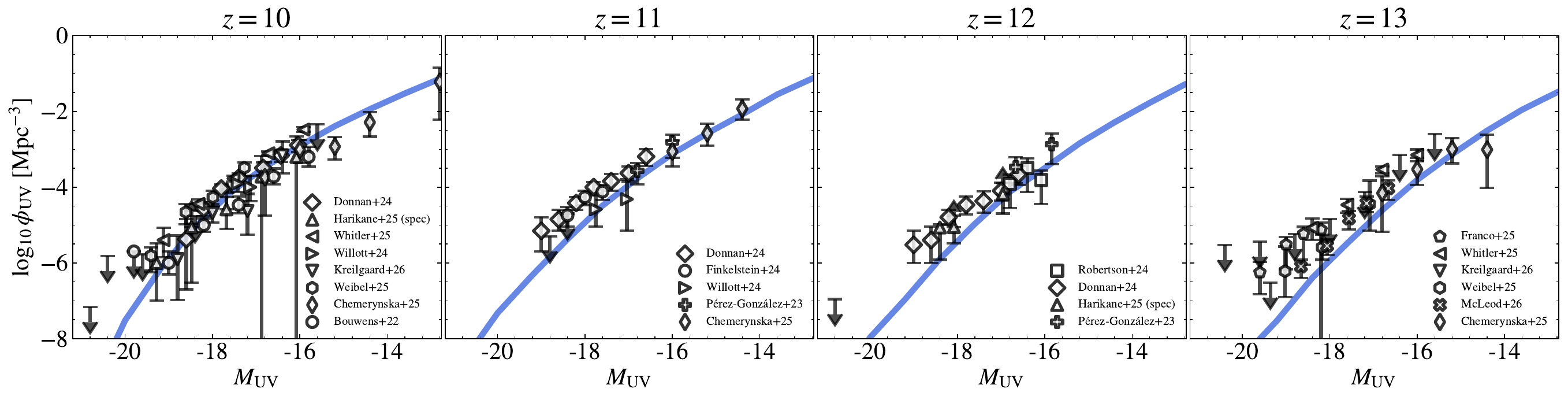}
    \caption{UV luminosity functions at $z=10$, 11, 12 and 13 (left to right). Blue curves show the predictions of the fiducial model with parameters calibrated at $z=6$ (Section~\ref{Sec:fiducial}), compared with JWST data \citep{Bouwens23, PerezGonzalez23, Donnan24, Finkelstein23, Robertson2024, Willott2024, Franco2025, Harikane2025, Weibel2025, Whitler25, Chemerynska2026, Kreilgaard2026, McLeod2026}. The model reproduces the UVLF up to $z=11$, but increasingly underpredicts the bright end toward higher redshifts.}
    \label{fig:uvlf_z10_13}
\end{figure*}
In Section~\ref{Sec:fiducial} we calibrated the fiducial model at $z=6$ to reproduce the observed UVLF \citep{Bouwens22lf} and the fractions of galaxies in star formation upturns and downturns \citep{Endsley25}. Here, we test how the model performs in reproducing UVLFs at higher redshift when we apply the same parameters, without further tuning.

Fig.~\ref{fig:uvlf_z10_13} shows the UVLFs predicted by the fiducial model, with the $z=6$ calibrated parameters compared with photometric and spectroscopic JWST constraints \citep{Bouwens23, PerezGonzalez23, Donnan24, Finkelstein23, Robertson2024, Willott2024, Franco2025, Harikane2025, Weibel2025, Whitler25, Chemerynska2026, Kreilgaard2026, McLeod2026}. 

At $z=10$ the model agrees well with the data across the full probed magnitude range. At $z=11$--12 the agreement remains good at the faint end, but the model starts underpredicting the data at the bright end, $M_{\rm UV}\lesssim-18$. The discrepancy between the fiducial model and the data grows with redshift. By $z=13$ the faint end of the model still agrees with the constraints from \citet{Chemerynska2026} but lies $\sim0.5$\,dex below those of \citet{Whitler25}, while the bright end falls short of the data by $\sim0.5$--$1$\,dex, anticipating the larger discrepancy at $z=14$ (Fig.~\ref{fig:z14models}).

\section{Star formation efficiency}\label{Appendix:sfe}

Star formation efficiency is a central quantity linking galaxy growth to dark matter halo assembly, and is widely used to interpret both stellar mass buildup and UV luminosity functions.
Our model describes star formation as stochastic fluctuations around a median efficiency. It is therefore useful to clarify how these fluctuations affect the effective star formation efficiency, particularly as their amplitude varies with halo mass.
A common definition of star formation efficiency is ${\rm SFE} = {\rm SFR}/(f_b \dot{M}_h)$. However, in the presence of bursty star formation histories, this definition is affected by stochastic variability in the star formation rate and its interpretation becomes non-trivial. In this section, we therefore quantify how burstiness modifies both the effective SFE and its scaling with halo mass.

In our framework, the star formation efficiency parameter $\epsilon_\star$ sets the normalization of the \textit{median} SFE at fixed halo mass, as galaxies undergo log-normal stochastic fluctuations around this median. Physically, this implies that star formation proceeds through intermittent bursts and downturns around a well-defined baseline efficiency.
Because of the lognormal nature of the fluctuations, the arithmetic \textit{mean} SFR is shifted above the median by a factor $e^{\sigma_{\rm \ln SFR}^2/2}$. This reflects the fact that high-SFR episodes contribute more to the total stellar mass formed than low-SFR phases suppress it. As a result, bursty star formation increases the cumulative stellar mass growth at fixed halo mass, leading to a higher effective star formation efficiency and UV luminosity.

\begin{figure}
    \centering
    \includegraphics[width=0.45\textwidth]{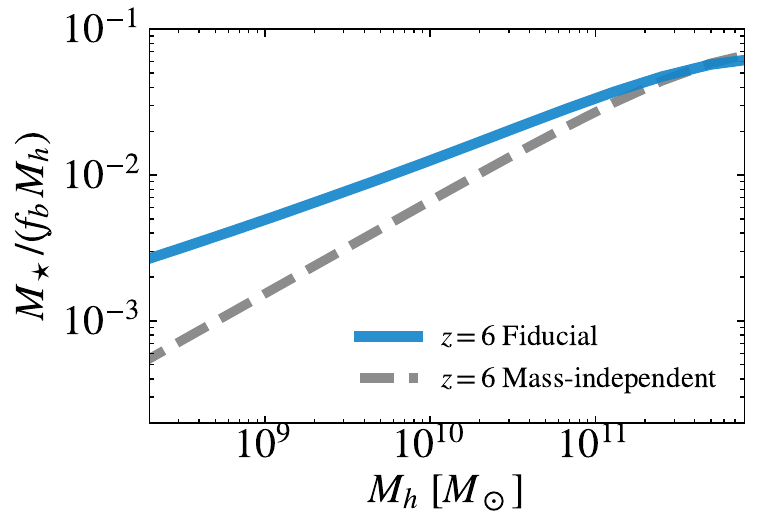}
    \caption{Integrated star formation efficiency, $M_\star / (f_b M_h)$, as a function of halo mass at $z=6$. The blue curve shows the fiducial mass-dependent burstiness model, while the grey dashed curve shows the mass-independent burstiness model. Mass-dependent burstiness produces a significantly shallower ${\rm SFE}$–$M_h$ relation compared to the mass-independent case.
    }
    \label{fig:sfe_z6}
\end{figure}

The integrated efficiency $M_\star / (f_b M_h)$ traces this cumulative effect of star formation over cosmic time. In Fig.~\ref{fig:sfe_z6}, we compare this quantity for our fiducial mass-dependent model and for a mass-independent burstiness model at $z=6$.

We find that mass-dependent burstiness leads to a significantly shallower ${\rm SFE}$–$M_h$ relation, similar to what found in FIREbox simulations by \cite{Feldmann2025}. Here, this arises because low-mass halos experience stronger stochastic fluctuations in log-SFR, which enhance their cumulative stellar mass growth through strong bursts more efficiently than in massive systems. In contrast, in the mass-independent model the burst-induced boost is uniform across halo masses, preserving a steeper ${\rm SFE}$–$M_h$ relation.

\bibliography{refer,codes,library_cm}
\bibliographystyle{aasjournalv7}  
 
\end{document}

%% file: include/def.tex
\def\be{\begin{equation}} 
\def\ee{\end{equation}} 
\def\ba{\begin{eqnarray}} 
\def\ea{\end{eqnarray}}

\def\msun{{\Msun}}

\newcommand{\NHI}{\ifmmode{N_\mathrm{HI}}\else $N_\mathrm{HI}$\xspace\fi}

\def\gsim{\lower.5ex\hbox{\gtsima}} 
\def\lsim{\lower.5ex\hbox{\ltsima}} \def\gtsima{$\; \buildrel > \over 
\sim \;$} \def\ltsima{$\; \buildrel < \over \sim \;$} \def\prosima{$\; 
\buildrel \propto \over \sim \;$} \def\gsim{\lower.5ex\hbox{\gtsima}} 
\def\lsim{\lower.5ex\hbox{\ltsima}} 
\def\simgt{\lower.5ex\hbox{\gtsima}} 
\def\simlt{\lower.5ex\hbox{\ltsima}} 
\def\simpr{\lower.5ex\hbox{\prosima}}   
  
 \def\gtsima{$\; \buildrel > \over \sim \;$} 
\def\ltsima{$\; \buildrel < \over \sim \;$} 
\def\gsim{\lower.5ex\hbox{\gtsima}} 
\def\lsim{\lower.5ex\hbox{\ltsima}} 
\def\simgt{\lower.5ex\hbox{\gtsima}} 
\def\simlt{\lower.5ex\hbox{\ltsima}} 
\def\simpr{\lower.5ex\hbox{\prosima}}

\newcommand{\Lya}{\ifmmode{\mathrm{Ly}\alpha}\else Ly$\alpha$\xspace\fi}

\def\msun{\,{\rm \Msun}}

\def\E3{{\cal E}_{\rm g}^{III}}

\def\r12{r_{1/2}} 
\def\x12{x_{1/2}} 
\def\v12{v_{1/2}}

\newcommand\code[1]{\textsc{\MakeLowercase{#1}}}

\def\nh2{n_{\rm H2}}
\def\fh2{f_{\rm H2}}

\def\angstrom{\textrm{A\kern -1.3ex\raisebox{0.6ex}{$^\circ$}}}

\def\msun{{\rm M}_{\odot}}

\def\msunyr{\msun\,{\rm yr}^{-1}}

